\documentclass[a4paper,11pt]{article}
\usepackage{jheppub} 
\usepackage[T1]{fontenc} 
\usepackage{graphicx}  
\usepackage{dcolumn,lineno}   
\usepackage{bm}        
\usepackage{amssymb}   
\usepackage{subfigure}
\usepackage{color}
\usepackage{xspace}
\usepackage{textpos}
\usepackage{epstopdf}
\usepackage{siunitx}
\newcolumntype{C}[1]{>{\centering}p{#1}}
\newcolumntype{L}[1]{>{\raggedright}p{#1}}
\newcolumntype{R}[1]{>{\raggedleft}p{#1}}
\newcolumntype{P}{@{$\ \pm\ $}}
\usepackage{tabularx}
\usepackage{dcolumn}
\usepackage{multirow}
\usepackage{verbatim}
\def \ee {e^{+}e^{-}}
\def \to {\rightarrow}
\def \pipi {\pi\pi}
\def \pp {\pi^{+}\pi^{-}}
\title{Measurement of $e^{+}e^{-}\rightarrow\omega\pi^{+}\pi^{-}$ cross section at $\sqrt s = $ 2.000 to 3.080 GeV}

\abstract{
A partial wave analysis on the process $e^{+}e^{-}\rightarrow\omega\pi^{+}\pi^{-}$ is performed using 647 pb$^{-1}$ of data sample collected by using the BESIII detector operating at the BEPCII storage ring at center-of-mass~(c.m.)~energies from 2.000 GeV to 3.080 GeV. The Born cross section of the $e^{+}e^{-}\rightarrow\omega\pi^{+}\pi^{-}$ process is measured, with precision improved by a factor of 3 compared to that of previous studies. A structure near 2.25 GeV is observed in the energy-dependent cross sections of $\ee\to\omega\pi^{+}\pi^{-}$ and $\omega\pi^{0}\pi^{0}$ with a statistical significance of 7.6$\sigma$, and its determined mass and width are 2232~$\pm$~19~$\pm$~27 MeV$/c^{2}$ and~\mbox{93~$\pm$~53~$\pm$~20 MeV}, respectively, where the first and second uncertainties are statistical and systematic,~respectively. By analyzing the cross sections of subprocesses $e^{+}e^{-}\rightarrow \omega f_{0}(500)$, $\omega f_{0}(980)$, $\omega f_{0}(1370)$, $\omega f_{2}(1270)$, and $b_{1}(1235)\pi$, a structure, with mass M~=~2200~$\pm$~11~$\pm$~17~MeV/$c^2$ and width $\Gamma$~=~74~$\pm$~20~$\pm$~24 MeV,~is observed with a combined statistical significance of 7.9$\sigma$. The measured resonance parameters will help to reveal the nature of vector states around 2.25 GeV. \\
}

\begin{document}
\maketitle
\flushbottom

\section{Introduction}
\vspace{-0.1cm}
The experiment of electron-positron collisions has long been a research topic in elementary particle physics wherein searches for exotic hadronic states have continuously received much attention. Some exciting observations in the $\ee$ experiments have been recently reported in
Refs.~\cite{zc3900,zc4020,beszc2,beszc3,beszc4,beszc5,beszc6,beszc7,beszc8,beszc9,beszcs,besZJY}. Meanwhile, measurements of the cross sections of exclusive processes are of particular interest for resolving the discrepancy of muon magnetic anomaly $a_\mu=(g_\mu-2)/2$ between experimental measurements and standard model (SM) predictions. The latest result by the Muon $g-2$ Experiment shows 3.3 standard deviation from the SM \mbox{prediction~\cite{muon2}}, implying a hint for the existence of new physics~\cite{standmodel}. However, to confirm this, further improvements in the precision for both experimental measurement and theoretical prediction are necessary. Specifically, the precision measurements on the cross sections of light hadron production in the $\ee$ annihilation are essential to reduce uncertainties of theoretical calculations on the hadronic vacuum polarization contribution and the muon anomaly $a_\mu$~\cite{theomuon,g2tohadron,g2tohadroncor}.

The BaBar experiment has studied the process $e^{+}e^{-}\rightarrow \omega\pi^+\pi^-$ using the initial state radiation (ISR) method~\cite{BABAR1}. An indication of an isoscalar resonance structure \mbox{near 2.25 GeV} was observed in its Born cross section line shape~\cite{BABAR0}. The systems $\omega f_{0}(500)$, $\omega f_{0}(980)$, $\omega f_{0}(1370)$, and $\omega f_{2}(1270)$ have isospin zero, which are useful to search for excited $\omega$ and $\phi$ states, and important to confirm the structure near 2.25 GeV observed through the BaBar experiment~\cite{BABAR0}. This structure may correspond to $\phi(2170)$, $\omega(2205)$, $\omega(2290)$, and $\omega(2330)$, whose properties are poorly known in the particle data group (PDG)~\cite{PDG2020}. Many interpretations have been proposed for $\phi(2170)$, including a traditional $s\bar{s}$ state~\cite{DingYanA,XWang,SSAfonin,CQPang,CGZhao,QLi}, an $s\bar{s}g$ hybrid~\cite{DingYan,JHoRBerg}, an $ss\bar{s}\bar{d}$ tetraquark state ~\cite{ZGWang,HXChen,NVDrenska,DengPing,HWKeLi,SSAgaev,RRDong,FXLiu}, a $\Lambda\bar{\Lambda}$ bound state~\cite{EKlempt,LZhao,CDeng,YBDong,YangChenLu}, and an ordinary resonant state of $\phi f_{0}(980)$ produced by the interactions between the final state particles~\cite{AMartinez,SGomez}. Recently,~the $\phi(2170)$ state was examined through the BESIII experiment in the processes of $e^{+}e^{-}\rightarrow K^{+}K^{-}$~\cite{LDBESIII}, $K^{+}K^{-}\pi^{0}\pi^{0}$~\cite{ZYTBESIII}, $\phi\eta^{\prime}$~\cite{SYQ2BESIII}, $\omega\eta$~\cite{GXLBESIII}, $K^{0}_{S}K^{0}_{L}$~\cite{MZXBESIII}, $\phi\eta$~\cite{PHETBESIII}, and $2(K^{+}K^{-})$,~$\phi K^{+}K^{-}$~\cite{FKPHBESIII}. The decay properties of $\omega(2205)$ as a candidate of $\omega(3D)$, $\omega(2290)$ and $\omega(2330)$ as candidates of $\omega(4S)$ states were discussed~\cite{w4s}. However, further experimental studies of the decay properties of these resonances are highly desired to reveal their natures.

In this paper, we report the measurement of Born cross sections for the $e^{+}e^{-}\rightarrow \omega\pi^+\pi^-$ process using the data collected in an energy scan at 19 c.m. energies ($\sqrt s$) from 2.000 GeV to 3.080 GeV with a total integrated luminosity of 647 pb$^{-1}$. The detailed values of c.m. energies and integrated luminosities of various data sets are presented in Section~\ref{CrsOfWpp}. Combined with the previous measurements of the $e^{+}e^{-}\rightarrow \omega\pi^0\pi^0$ process~\cite{Bes3Wpp}, the cross sections of the $\ee\to\omega \pi\pi$ process are obtained. In addition, the cross sections of the subprocesses via some intermediate states (e.g., $e^{+}e^{-}\rightarrow \omega f_{0}(500)$,~$\omega f_{0}(980)$,~$\omega f_{0}(1370)$,~$\omega f_{2}(1270)$, and $b_{1}(1235)\pi$) are determined.

\section{BESIII detector and Monte Carlo simulation}
The BESIII detector~\cite{bes3} records symmetric $e^+e^-$ collisions
provided by the BEPCII storage ring~\cite{bepc2}, which operates in the c.m. energy range within 2.000--4.950~GeV. BESIII has collected large data samples in this energy region~\cite{Ablikim:2019hff}. The cylindrical core of the BESIII detector covers 93\% of the full solid angle and comprises a helium-based multilayer drift chamber~(MDC), a plastic scintillator time-of-flight system~(TOF), and a CsI(Tl) electromagnetic calorimeter~(EMC), which are all enclosed in a superconducting solenoidal magnet providing a 1.0~T magnetic field. The solenoid is supported by an octagonal flux-return yoke with resistive plate counter muon identification modules interleaved with steel.
The charged-particle momentum resolution at $1~{\rm GeV}/c$ is $0.5\%$, and the specific ionization energy loss $dE/dx$ resolution is $6\%$ for electrons
from Bhabha scattering. The EMC measures photon energies with a resolution of $2.5\%$ ($5\%$) at $1$~GeV in the barrel (end cap)
region. The time resolution in the TOF barrel region is 68~ps, while that in the end cap region is 110~ps.

Monte-Carlo (MC) events, including the geometric description of the BESIII detector and the detector response, are produced using GEANT4-based~\cite{geant4} offline software BOSS~\cite{bes3}. Two million of inclusive MC events, $\ee\to\text{hadrons}$, are used to estimate the background contamination. They are generated using a hybrid generator, which integrates PHOKHARA~\cite{phokhara,hjsgen}, ConExc~\cite{hybridgen,exlusgen2}, and LUARLW~\cite{lundargen} models. The PHOKHARA model generates 10 well parameterized and established exclusive channels. The ConExc model simulates a total of 47 exclusive processes according to a homogeneous and isotropic phase space population and then reproduces the measured line shapes of the absolute cross section. The remaining unknown decays of virtual photon are modeled by the LUARLW model.

Signal MC events for the $e^+e^-\to\omega\pi^+\pi^-$ process with subsequent decays $\omega\to \pi^{+}\pi^{-}\pi^{0}$ and $\pi^{0}\to \gamma\gamma$ are generated by using the amplitude model with parameters fixed to the helicity amplitude analysis results, and the significant intermediate states are included with the ISR effects.

\section{Event selection and background analysis}
A data sample of the $e^{+}e^{-}\rightarrow \pi^{+}\pi^{-}\pi^{+}\pi^{-}\pi^{0}$ process is selected by requiring the net charge of four charged tracks to be zero and at least two photons. Charged tracks detected in the MDC are required to be within a polar angle ($\theta$) range of $|\!\cos\theta|$<0.93, where $\theta$ is defined with respect to the z-axis, which is the
symmetry axis of the MDC. For charged tracks, the distance of the closest approach to the interaction point (IP) must be less than 10\,cm along the z-axis of the MDC, $|V_{z}|$, and less than 1\,cm in the transverse plane, $|V_{xy}|$. Photon candidates are identified using showers in the EMC.  The deposited energy of each shower is more than 25~MeV in the barrel region ($|\!\cos\theta|$<0.80) and more than 50~MeV in the end cap region (0.86<$|\!\cos\theta|$<0.92). To exclude showers that originate from charged tracks, the angle between the position of each shower in the EMC and the closest extrapolated charged track should be greater than 10 degrees. To suppress electronic noise and showers unrelated to the event, the difference between the EMC time and event start time must be \mbox{within (0, 700) ns}.

To improve the mass resolution and suppress the background contribution, a four-constraint (4\rm{C}) kinematic fit  imposing four-momentum conservation is performed under the hypothesis of the  $e^+e^-\to2(\pi^+\pi^-)\gamma\gamma$ process. When more than two photons exist, the  combination with minimum $\chi^{2}_{4\rm{C}}$ is retained for further analysis. Based on an optimization of the signal-to-noise ratio $\frac{N_s}{\sqrt{N_s+N_b}}$ for the requirement on $\chi^{2}_{4\rm{C}}$, where $N_s$ and $N_s+N_b$ are the numbers of events in the $\omega$ signal region for signal MC sample and data, respectively,
candidate events with $\chi^{2}_{4\rm{C}}<40$ are accepted for further analysis. The $\omega\pi^{+}\pi^{-}$ events are reconstructed by selecting the $\omega$ candidates from all the four $\pi^{+}\pi^{-}\pi^{0}$ combinations with minimal mass difference from the $\omega$ known mass~\cite{PDG2020}. The $\pi^0$ candidate is reconstructed by the $\gamma\gamma$ combination with a mass window of 0.12<$M_{\gamma\gamma}$<0.15 GeV/$c^{2}$, and the $\omega$ signal region is defined with a requirement of 0.758<$M_{\pi^+\pi^-\pi^0}$<0.808 GeV/$c^{2}$. To estimate the background, the $\omega$ sidebands are defined as [0.733 , 0.758] GeV/$c^{2}$ and [0.808, 0.833] GeV/$c^{2}$. The $e^{+}e^{-}\rightarrow\pi^{+}\pi^{-}\pi^{+}\pi^{-}\pi^{0}$ process is found to be the dominant background source by analyzing the inclusive MC sample~\cite{XYZTopAna}. No peaking background is found in the signal region.

The selection of the 3$\pi$ combination for $\omega$ candidate may lead to a wrong $\pi^+\pi^-\pi^0$ combination. The ratio due to the mis-combination is examined by matching the MC truth to the reconstructed final state particles using a sample of $0.5$ million MC events for the $e^{+}e^{-}\to 2(\pi^{+}\pi^{-})\pi^{0}$ process. When a charged pion is assigned to be from the wrong $\omega$ candidate, an event will be marked as "mis-combination". The ratio of the number of mis-combination $\omega$  to the total number of events is about $1.0\%$ on average. The mis-combination yields will be subtracted from the $\omega$ candidate events when calculating the $e^+e^-\to\omega\pi^+\pi^-$ Born cross section.

\section{Amplitude analysis}
\subsection{Kinematic variable}
\begin{figure}[htbp]
\begin{center}
\includegraphics[width=\textwidth,angle=0]{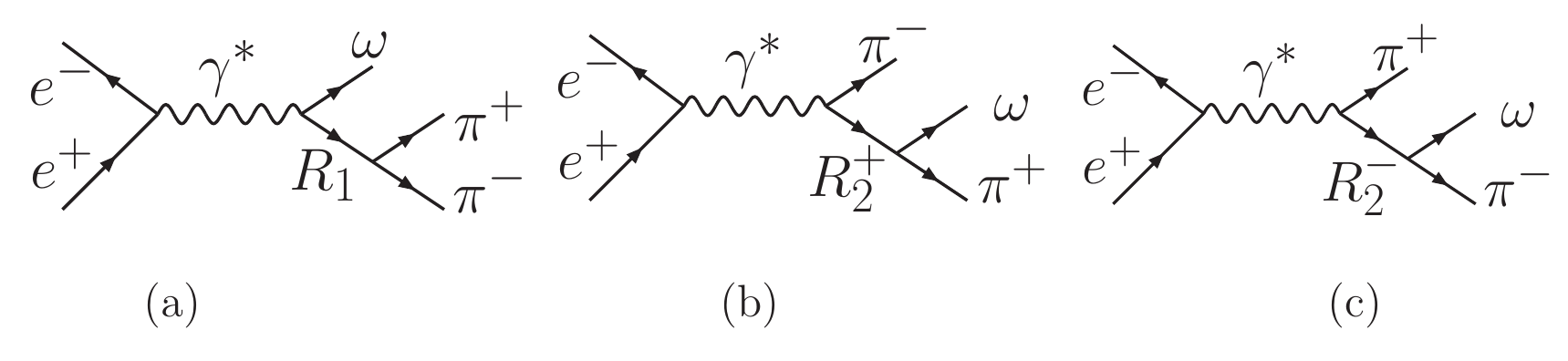}
\put(-280,90){\rotatebox{90}} \caption{Feynman diagrams of the quasi-two body processes for $\ee\to\omega\pp$.} \label{bornprocess}
\end{center}
\end{figure}

\begin{figure}[htbp]
\begin{center}
\includegraphics[width=\textwidth,angle=0]{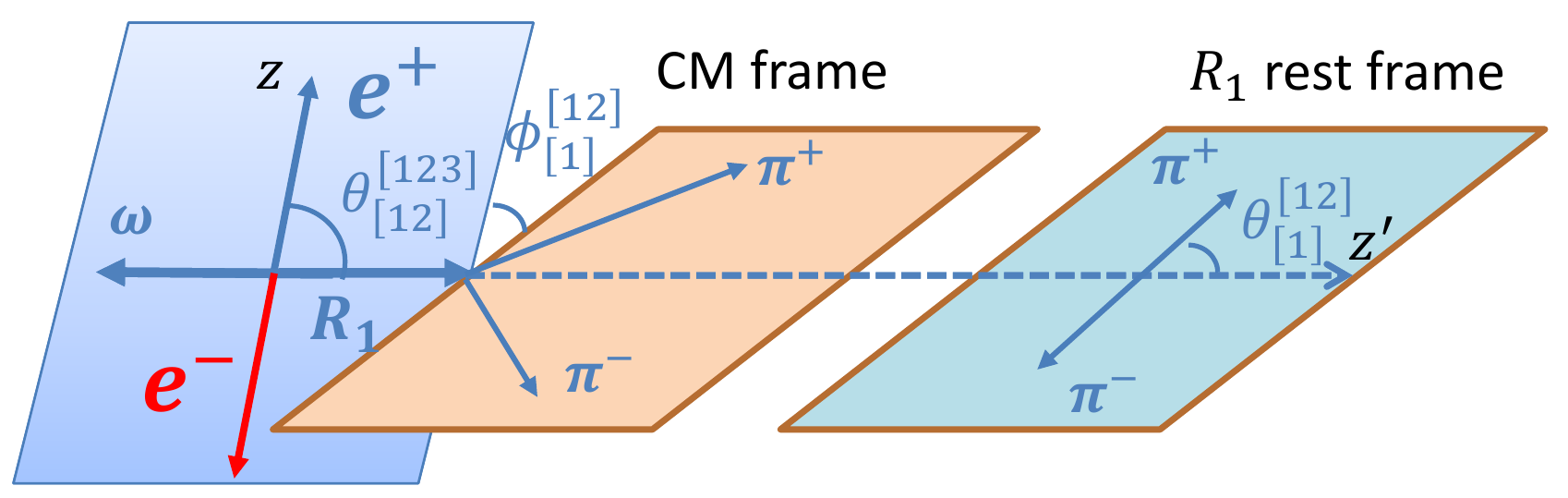}
\put(-280,90){\rotatebox{90}} \caption{Helicity rotation angles defined for the process $\ee\to R_1\omega,~R_1\to  \pi^+\pi^-$.} \label{helsys}
\end{center}
\end{figure}

The $e^+e^-\rightarrow\omega\pp$ events selected with the kinematic constraints are produced from the $\ee$ annihilation into a virtual photon, which then transformed to light quarks, followed by hadronization into the $\pi^+(p_1)\pi^-(p_2)\omega(p_3)$ final sates, where $p_i~(i=1,2,3)$ denote the momenta reconstructed for the associated particles. The $e^+e^-\rightarrow\omega\pi^+\pi^-$ process may be from the quasi-two body process of $\ee\to R_1\omega, ~R_1\to\pi^+\pi^-$~(a); $\ee\to \pi^-R_2^+$, \mbox{$~R_2^+\to \pi^+\omega$~(b)} and $\ee\to \pi^+R_2^-, ~R_2^-\to \pi^-\omega$ (c), with Feynman diagrams shown in Fig.~\ref{bornprocess}, where $R_1$ and $R_2^\pm$ denote the intermediate states.
In the amplitude analysis, the intensity for each process is described with the helicity amplitude, which is constructed as the amplitude production of two sequential decays in the helicity frame. Taking process (a) as an example, Fig.~\ref{helsys} shows the definition of helicity rotation angles, where the $R_1$ polar angle, $\theta_{[12]}^{[123]}$, is defined as the angle spanned between the $R_1$ momentum and positron moving direction, and the azimuthal angle $\phi_{[12]}^{[123]}$ is the angle between the $R_1$ production plane and its decay plane. For the $R_1\to\pp$ decay, the azimuthal angle, $\phi_{[1]}^{[12]}$, is defined as the angle between $R_1$ production plane and its decay plane, formed by the $\pp$ momentum. After boosting the two pion momenta to the $R_1$ rest frame, they remain in the same decay plane. The polar angle $\theta_{[1]}^{[12]}$ for $\pi^+$ is defined as the angle between the $\pp$ and $R_1$ momentum in the $R_1$ rest frame. Helicity angles for processes (b) and (c) are defined in the same manner. Table \ref{tabAngAmp} lists the definitions of helicity angles and amplitudes for the three processes.

\begin{table}[htbp]
\begin{center}
\caption{Definitions of helicity angles and amplitudes for sequential processes (a), (b), and (c). Here, $\lambda_i$ and $m$ denote the helicity values for individual particle and the spin $z$ projection of virtual photon in electron-positron annihilation.~\label{tabAngAmp} }
\vspace{0.3cm}
\begin{tabular}{lll}
\hline\hline
 Processes & Helicity angles & Amplitudes\\\hline
$\ee\to\gamma^*(m)\to R_1(\lambda_R)\omega(\lambda_3)$ & $\theta_{[12]}^{[123]},\phi_{[12]}^{[123]}$ & $F^{\gamma^*}_{\lambda_R,\lambda_3}$\\
$R_1\to \pi^+\pi^-$ & $\theta_{[1]}^{[12]},\phi_{[1]}^{[12]}$ & $F^{R_1}_{0,0}$\\\hline
$\ee\to\gamma^*(m)\to R_2^+ (\lambda_{+})\pi^-$ & $\theta_{[13]}^{[123]},\phi_{[13]}^{[123]}$ & $F^{\gamma^*}_{\lambda_{+},0}$\\
$R_2^+\to\omega(\lambda_3^{'})\pi^+$ & $\theta_{[3]}^{[13]},\phi_{[3]}^{[13]}$ & $F^{R_2^+}_{\lambda_3^{'},0}$\\\hline
$\ee\to\gamma^*(m)\to R_2^- (\lambda_{-})\pi^+$ & $\theta_{[23]}^{[123]},\phi_{[23]}^{[123]}$ & $F^{\gamma^*}_{\lambda_{-},0}$\\
$ R_2^- \to\omega(\lambda_3^{''})\pi^-$ & $\theta_{[3]}^{[23]},\phi_{[3]}^{[23]}$ & $F^{ R_2^- }_{\lambda_3^{''},0}$\\\hline\hline
\end{tabular}
\end{center}
\end{table}

\subsection{Helicity amplitude}

The amplitude for process (a) reads
\begin{equation}
A_1(m,\lambda_3)=\sum_{\lambda_R} F^{\gamma^*}_{\lambda_R,\lambda_3}D^{1*}_{m,\lambda_R-\lambda_3}(\phi_{[12]}^{[123]},\theta_{[12]}^{[123]},0)BW(m_{12})F^{R_1}_{0,0}D^{J*}_{\lambda_R,0}(\phi_{[1]}^{[12]},\theta_{[1]}^{[12]},0),
\end{equation}
where $D^J_{m,\lambda}(\phi,\theta,0)$ is the Wigner-$D$ function, $J$ is the spin of resonance $R_1$, and $BW$ denotes the Breit-Wigner function.

The amplitude for process (b) reads
\begin{eqnarray}
A_2(m,\lambda_3)&=&\sum_{\lambda_+,\lambda_3'} F^{\gamma^*}_{\lambda_+,0}D^{1*}_{m,\lambda_+}(\phi_{[13]}^{[123]},\theta_{[13]}^{[123]},0)
BW(m_{13})F^{R_2^+}_{\lambda_3',0}D^{J*}_{\lambda_+,\lambda_3'}(\phi_{[3]}^{[13]},\theta_{[3]}^{[13]},0)\nonumber\\
&\times&D^{1}_{\lambda_3',\lambda_3}(\phi_3',\theta_3',0),
\end{eqnarray}
where $J$ is the spin of $R_2^+$. Since the $\omega$ helicity defined in this process is different from that defined in process (a), a rotation should be performed to align the $\omega$ helicity to coincide with that in process (a) by the angle $(\theta_3',\phi_3')$. This issue has been addressed in the \mbox{analyses~\cite{lhcb,belle}} and proved in Ref.~\cite{pingrg}.

The amplitude for process (c) reads
\begin{eqnarray}
A_3(m,\lambda_3)&=&\sum_{\lambda_-,\lambda_3''} F^{\gamma^*}_{\lambda_-,0}D^{1*}_{m,\lambda_-}(\phi_{[23]}^{[123]},\theta_{[23]}^{[123]},0)
BW(m_{23})F^{ R_2^- }_{\lambda_3'',0}D^{J*}_{\lambda_-,\lambda_3''}(\phi_{[3]}^{[23]},\theta_{[3]}^{[23]},0)\nonumber\\
&\times&D^{1}_{\lambda_3'',\lambda_3}(\phi_3'',\theta_3'',0),
\end{eqnarray}
where $J$ is the spin of $R_2^-$. The Wigner function $D^{J}_{\lambda_3',\lambda_3}(\phi_3'',\theta_3'',0)$ is used to align the $\omega$ helicity to coincide with that defined in process $(a)$.

The non-resonant three-body process $\ee\to\omega\pp$ is also considered, and its amplitude is written as~\cite{berman,chung2,chung20,chung21}:
\begin{eqnarray}
A_4(m,\lambda_3)&=&\sum_{\mu} F_{\mu,\lambda_3}D^{1*}_{m,\mu}(\alpha,\beta,\gamma),
\end{eqnarray}
where $\mu$ is the $z$-component of spin $J$ in the helicity system; $m(\lambda_3)$ is the helicity value for $\gamma^*(\omega)$; $\alpha,\beta,\gamma$ are the Euler angles as defined in Refs.~\cite{berman,chung2,chung20,chung21}; and $F_{\mu,\lambda_3}$ is the helicity amplitude. The process conserves the parity and thus leads to $F_{\pm,\lambda_3}=-F_{\pm,-\lambda_3}$ and $F_{0,\lambda_3}=F_{0,-\lambda_3}$.

To match the covariant tensor amplitude, we expand the helicity amplitude in terms of the partial waves for
the two-body process in the $LS$-coupling scheme~\cite{chung2,chung20,chung21}. It follows
\begin{eqnarray}\label{chung_forma}
F^J_{\lambda,\nu} &=&\sum_{lS}\left ({2l+1\over 2J+1}\right )^{1/2}\langle l0S\delta|J\delta\rangle
\langle s\lambda\sigma-\nu|S\delta\rangle g_{lS}r^l{B_l(r)\over B_l(r_0)},
\end{eqnarray}
for a spin-$J$ particle decay $J\rightarrow s+\sigma$, and $\lambda$ and $\nu$ are the helicities of two final-state particles $s$ and $\sigma$ with $\delta=\lambda-\nu$. The symbol $g_{lS}$ is a coupling constant, $S$ is the total intrinsic spin ${\bf S =s+\sigma}$, and $l$ is the orbital angular momentum, $r=|{\bf r}|$, where ${\bf r}$ is the relative momentum between the two daughter particles in their mother rest frame and ${\bf r}_0$ corresponds to the invariant mass of the resonance which is equal to its nominal mass.
The Blatt-Weisskopf factor $B_l(r)$~\cite{chung2,chung20,chung21} up to $l=4$ takes the form
\begin{eqnarray}\label{}
B_0(r)/B_0(r_0)&=&1,\nonumber\\
B_1(r)/B_1(r_0)&=&\frac{\sqrt{1+(dr_0)^{2}}}{\sqrt{1+(dr)^{2}}},\nonumber\\
B_2(r)/B_2(r_0)&=&\frac{\sqrt{9+3(dr_0)^{2}+(dr_0)^{4}}}{\sqrt{9+3(dr)^{2}+(dr)^{4}}},\\
B_3(r)/B_3(r_0)&=&\frac{\sqrt{225+45(dr_0)^{2}+6(dr_0)^{4}+(dr_0)^{6}}}{\sqrt{225+45(dr)^{2}+6(dr)^{4}+(dr)^{6}}},\nonumber\\
B_4(r)/B_4(r_0)&=&\frac{\sqrt{11025+1575(dr_0)^{2}+135(dr_0)^{4}+10(dr_0)^{6}+(dr_0)^{8}}}{\sqrt{11025+1575(dr)^{2}+135(dr)^{4}+10(dr)^{6}+(dr)^{8}}},\nonumber
\end{eqnarray}
where parameter $d$ is constant fixed to 3 GeV$^{-1}$~\cite{lhcb}.

The differential cross section is given by
\begin{equation}\label{xsection}
{\rm d}\sigma={1\over 2}\sum_{m,\lambda_3}\Omega(\lambda_3)\left|\sum_{i=1}^3A_i(m,\lambda_3)\right|^2{\rm d\Phi},
\end{equation}
where $m=\pm1$ is due to the virtual photon produced from unpolarized $\ee$ annihilations, $\lambda_3$ is the $\omega$ helicity value, and ${\rm d}\Phi$ is the element of a standard three-body phase space. The variable $\Omega(\lambda_3)(=|{\bf\varepsilon}(\lambda_3)\cdot{\bf (q_1\times q_2)}|^2$) is the $\omega$ decay matrix into the $\pp\pi^{0}$ final states, where $\varepsilon$ is the $\omega$ polarization vector and ${\bf q_1(q_2)}$ is the momentum vector for $\pi^+(\pi^-)$ from the $\omega$ decay. Here, we factor out the Breit-Wigner function describing the $\omega$ line shape into the MC integration when applying the amplitude model fit to the data events.
\subsection{Simultaneous fit}
The relative magnitudes and phases for coupling constants are determined
by an unbinned maximum likelihood fit. The joint probability for observing $N$ events in one data set is
\begin{equation}
\mathcal{L}=\prod_{i=1}^N P_i(p_1,p_2,p_3,p_4,p_5),
\end{equation}
where $p_j~(j=1,2,...,5)$ denote the four-vector momenta of the final states and $P_i$ is a probability to produce event $i$ with the momenta of the final states. The normalized $P_i$ is calculated from the differential cross section
\begin{equation}
P_i={({\rm d}\sigma /{\rm d}\Phi)_i \over \sigma_{\rm MC}}.
\end{equation}
Here, the normalization factor $\sigma_{\rm MC}$ is calculated using a signal MC
sample with $N_{\rm MC}$ accepted events, which are generated with a uniform distribution in phase
space and subjected to the detector simulation, and then are
passed through the same event selection criteria as applied to the
data events. With a signal MC sample of two million events, the
$\sigma_{\rm MC}$ is approximately evaluated as
\begin{equation}
\sigma_{\rm MC}\approx{1\over N_{\rm MC}}\sum_{i=1}^{N_{\rm MC}}\left({{\rm d}\sigma\over {\rm d}\Phi}\right)_i.
\end{equation}
For technical reasons, rather than maximizing $\mathcal{L}$, the object function, $S=-\ln\mathcal{L}$, is minimized using MINUIT~\cite{minuit}. To subtract the background events, the object function is replaced with

\begin{equation}
S=-\ln\mathcal{L} = -(\ln\mathcal{L}_\textrm{data}-\ln\mathcal{L}_\textrm{bg}),
\end{equation}
where $\mathcal{L}_\text{data}$ and $\mathcal{L}_\text{bg}$ are the joint probability densities for data and background, respectively. The background events are obtained by the normalized $\omega$ sidebands.

We perform a simultaneous fit to an ensemble of the $\omega\pp$ events with different c.m. energies, and their amplitudes sharing same coupling constants describing the intermediate state processes. The total object function is obtained as the summation of individual ones:
\begin{equation}
S' = -\sum_{j=1}^N \ln\mathcal{L}_j,
\end{equation}
for the total $N$ sets of data events.

After the parameters are determined in the fit, the signal yield of a given subprocess can be estimated by scaling its cross section ratio $r_i$ to the number of net events
\begin{equation}\label{yieldsFormula}
N_i=r_i(N_\textrm{\scriptsize{obs}}-N_\textrm{\scriptsize{bg}}),\textrm{~with~}r_i={\sigma_i\over \sigma_\textrm{\scriptsize{tot}}},
\end{equation}
where $\sigma_i$ is the cross section for the $i$-th subprocess as defined in Eq.(\ref{xsection}), $\sigma_\textrm{tot}$ is the total cross section, $N_\textrm{obs}$ is the number of observed events, and $N_\textrm{bg}$ is the number of background events.

The statistical uncertainty, $\delta N_i$, associated with the signal yield $N_i$ is estimated according to the error propagation formula using the covariance matrix $V$, obtained from the fit:
\begin{equation}\label{staterr}
\delta N_{i}^{2} = \sum_{m=1}^{N_\textrm{\scriptsize{pars}}}\sum_{n=1}^{N_\textrm{\scriptsize{pars}}}\left({\partial N_i\over \partial X_m}{\partial N_i\over \partial X_n}\right)_{\bf{X}={\bf \mu}}V_{mn}({\bf X}),
\end{equation}
where ${\bf X}$ is a vector containing parameters and ${\bf \mu}$ contains the fitted values for all parameters. The sum runs over all $N_\textrm{pars}$ parameters.

\subsection{Intermediate states and significance check}
  In terms of the $\pp$ mass spectrum shown in Fig.~\ref{fig:A} for 12 of the 19 samples, the $f_0(980)$ signals are significantly observed in the data events at c.m. energies below 2.3094 GeV. In the low $\pp$ energy region, $f_0(500)$ resonance may have some contributions. In the high energy region, $f_0(1370)$ and $f_2(1270)$ resonances are included in the amplitude model. The $f_0(980)$ line shape is parameterized with the Flatt\^e~formula:
\begin{equation}\label{besiia,besiib}
BW(s)={1\over s-m_{0}^2+i(g_1\rho_{\pi\pi}(s)+g_2\rho_{K\bar K}(s))},
\end{equation}
where $\rho(s)=2k/\sqrt s$; $k$ is the c.m. momentum of the $\pi$ or $K$ in the resonance rest frame; $g_1$ and $\frac{g_2}{g_1}$ are fixed to the measured values of $0.138\pm0.010 ~$GeV$^2$ and $4.45\pm0.25$, respectively~\cite{besiia,besiib}; and $m_{0}$ is the mass of the resonance $f_{0}(980)$ ($m_{0}$=0.990 GeV/$c^{2}$).

For the $f_0(500)$ line shape, many types of energy-dependence-width parameterizations exist in the literature~\cite{besiia,besiib}, among which we choose the E791 parameterizations (used by the E791 Collaboration) in the nominal fit:
\begin{equation}
BW(s)={1\over s-m_0^2+i\sqrt{s} \Gamma},~\text{with~}
\Gamma=\sqrt{1-{4m_\pi^2\over s }}\Gamma_0,
\end{equation}
where $m_0$ and $\Gamma_0$ are the $f_0(500)$ mass and width, respectively.

For the other resonances (e.g., $f_{0}(1370)$,~$f_{2}(1270)$,~$b_{1}(1235)$,~$\rho(1450)^\pm$ and $\rho(1570)^\pm$), the width is considered constant, and the line shape is described with the Breit-Weigner function, in which their masses and widths are fixed to the PDG values, as given in Table~\ref{sigtable}.

The partial wave analysis (PWA) fit procedure begins by including all possible intermediate states in the PDG that match $J^{PC}$ conservation in the subsequent two-body decay.
Then we examine the statistical significance of the individual amplitudes. Amplitudes with statistical significance less then 5$\sigma$ are dropped. This procedure is repeated until a baseline solution is obtained with only amplitudes having a statistical significance greater than 5$\sigma$.

The 12 samples were divided into two groups, with group A including the energy points of $\sqrt s=2.0000,~2.1000,~2.1250,~2.1750,~2.2000,~2.2324$ GeV, and group B including $\sqrt s=2.3094, ~2.3864$,~$2.3960,~2.6444,~2.6464,~2.9000$ GeV according to the $\pi^{+}\pi^{-}$ invariant mass distributions at various energy points.

The statistical significance of each amplitude is evaluated by incorporating the changes in likelihood and number of degrees of
freedom with and without the corresponding amplitude being included in the simultaneous fit. The significance for each intermediate state is listed in Table~\ref{sigtable}. We take the subprocess with a significance greater than 5$\sigma$ in both group A and group B data sets to be the baseline solutions, including the \mbox{$e^+e^-\to \omega f_0(500),$}~$\omega f_0(980),\\~\omega f_0(1370),~\omega f_2(1270)$ and $b_1^{\mp}(1235)\pi^{\pm}$ processes.

\vspace{-0.6cm}
\begin{table*}[htbp]
\setlength{\belowcaptionskip}{10pt}
\caption{Masses and widths of the intermediate states and statistical significances of individual intermediate processes in group A and group B data sets.\label{sigtable}}\centering{
\vspace{0.3cm}
\begin{tabular}{ccccc}
\hline\hline
Resonance & Mass (GeV/c$^{2}$) & Width (GeV) & Group A & Group B\\\hline
$f_0(500)$ & 0.507 (0.400$\sim$0.550) & 0.475 (0.400$\sim$0.700) & 8.4$\sigma$  & 14.0$\sigma$\\
$f_0(980)$ &0.990 $\pm$ 0.020 & ---&15.0$\sigma$&11.8$\sigma$\\
$f_0(1370)$ &1.350 $\pm$ 0.050&0.200 $\pm$ 0.500&12.6$\sigma$&9.1$\sigma$\\
$f_2(1270)$ &1.2755 $\pm$ 0.0008&0.1867 $\pm$ 0.0022&12.6$\sigma$&11.0$\sigma$\\
$b_1^\pm(1235)$ &1.2295 $\pm$ 0.0032&0.142 $\pm$ 0.009&11.1$\sigma$&19.0$\sigma$\\
$\rho(1450)^\pm$&1.465 $\pm$ 0.025 &0.400 $\pm$ 0.060&4.4$\sigma$&8.3$\sigma$\\
$\rho(1570)^\pm$&1.570 $\pm$ 0.070&0.144 $\pm$ 0.090&6.1$\sigma$&4.3$\sigma$\\
\hline\hline
\end{tabular}
}
\end{table*}
\vspace{-1cm}
\begin{table*}
\begin{center}
\caption{Signal yields of the subprocesses at various selected energy points, where the uncertainties are only statistical. The last column is the non-resonance part.}
\vspace{0.3cm}
\begin{small}
\begin{tabular}{L{1.cm}R{1.cm}PR{0.4cm}R{0.8cm}PR{0.4cm}R{1.cm}PR{0.4cm}R{1.cm}PR{0.4cm}R{0.8cm}PR{0.4cm}R{0.6cm}PR{0.4cm}r} \hline\hline
 $\sqrt{s}$~(GeV) & \multicolumn{2}{c}{$\omega f_0(500)$} & \multicolumn{2}{c}{$\omega f_0(980)$} & \multicolumn{2}{c}{$\omega f_0(1370)$} &\multicolumn{2}{c} {$\omega f_2(1270)$}  &\multicolumn{2}{c} {$b_1^{\mp}(1235)\pi^{\pm}$}&\multicolumn{2}{c} {$\omega\pi^{+}\pi^{-}$}\\ \hline
  2.0000& 58.6&23.6& 446.4&42.4& 200.4&50.3& 167.2&26.8& 27.8&8.9& 1.1&1.1& \\
  2.1000& 45.1&19.9& 578.4&46.7 & 197.6&46.3& 130.0&25.1& 9.3&6.6&10.9&10.1& \\
  2.1250& 720.2&76.9& 2281.&104.8& 787.2&117.8& 759.5&85.6& 316.0&41.1&24.6&5.1& \\
  2.1750& 130.3&23.7& 242.9&29.7& 59.0&24.5&  57.1&20.3 & 55.4&11.5&15.4&8.1&  \\
  2.2000& 98.5&20.7& 262.9&31.1& 305.7&47.0& 146.7&30.5& 13.3&7.8& 13.0&10.3& \\
  2.2324& 100.5&15.7& 242.8&25.7& 187.7&33.9& 50.5&14.6& 23.9&8.9&64.0&8.8& \\
  2.3094& 205.4&41.7& 63.0&24.8& 32.1&26.9& 145.8&33.8& 17.6&9.9 &12.1&6.3&\\
  2.3864& 115.1&21.4& 35.7&15.4& 72.7&30.7& 167.3&32.5 & 26.2&8.8& 43.0&7.3& \\
  2.3960& 441.1&42.3& 70.8&24.3& 399.2&62.4& 115.6&33.8& 133.0&27.5&15.4&10.7& \\
  2.6444& 97.6&20.1& 42.8&14.8& 49.2&22.3& 65.6&19.8& 42.9&11.4&97.0&22.2& \\
  2.6464& 104.0&15.6& 85.6&21.2& 100.7&25.8& 69.2&14.2& 75.6&10.3&18.3&12.4& \\
  2.9000& 194.0&24.7& 35.0&17.2& 47.4&25.4& 58.5&19.1& 112.0&13.2&32.5&6.7&
  \\ \hline \hline
\end{tabular}
\label{tabsigyields}
\end{small}
\end{center}
\end{table*}

\vspace{1.5cm}
\subsection{Fit results}
\begin{figure*}[htbp]
\begin{center}
\includegraphics[width=\textwidth]{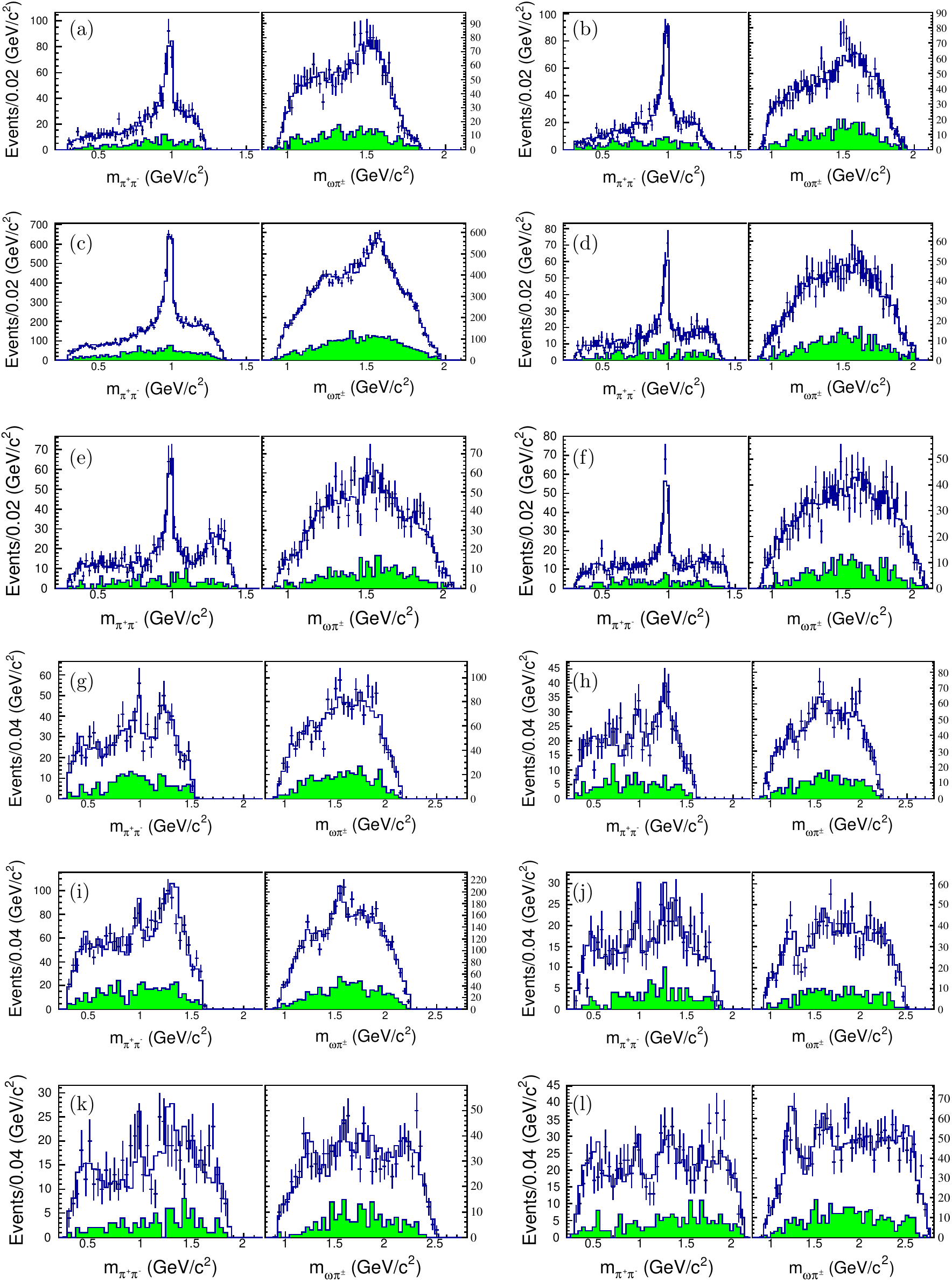}
\caption{Projections of the fit results for mass spectra $m_{\pp}$ and $m_{\omega\pi^\pm}$ for selected energy points. Dots with error bars are the data, histograms are the projection of the fit results, and shaded histograms (green) are the background estimated by the $\omega$ mass sideband. The subplots are for data sets $\sqrt{s}=$~2.0000~(a),~2.1000~(b),~2.1250~(c), ~2.1750~(d),~2.2000~(e),~2.2324~(f),~2.3094~(g),~2.3864~(h),~2.3960~(i),~2.6444~(j),~2.6464~(k), and~2.9000~(l) GeV.}
\label{fig:A}
\end{center}
\end{figure*}

\begin{figure}[htbp]
\begin{center}
\subfigure{
        \includegraphics[scale=0.5]{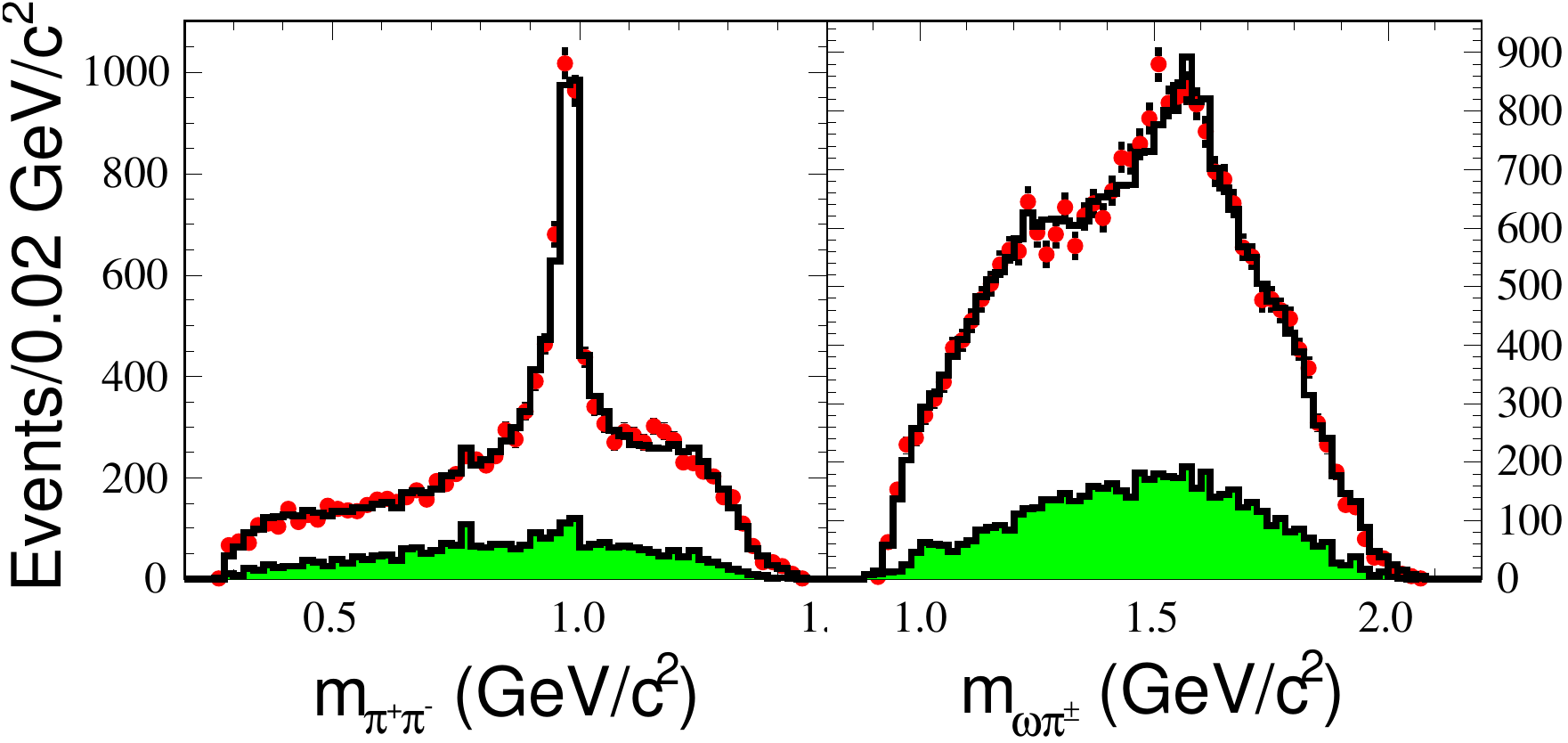}
	}
\caption{Projections of the fit results for mass spectra $m_{\pp}$ and $m_{\omega\pi^\pm}$ for group A data sets. Dots with error bars are the data, histograms are the projection of the total fit results, and shaded histograms (green) are the background estimated by the $\omega$ mass sideband.}
\label{fig:ToTA}
\end{center}
\end{figure}

\begin{figure}[htbp]
\begin{center}
\subfigure{
        \includegraphics[scale=0.5]{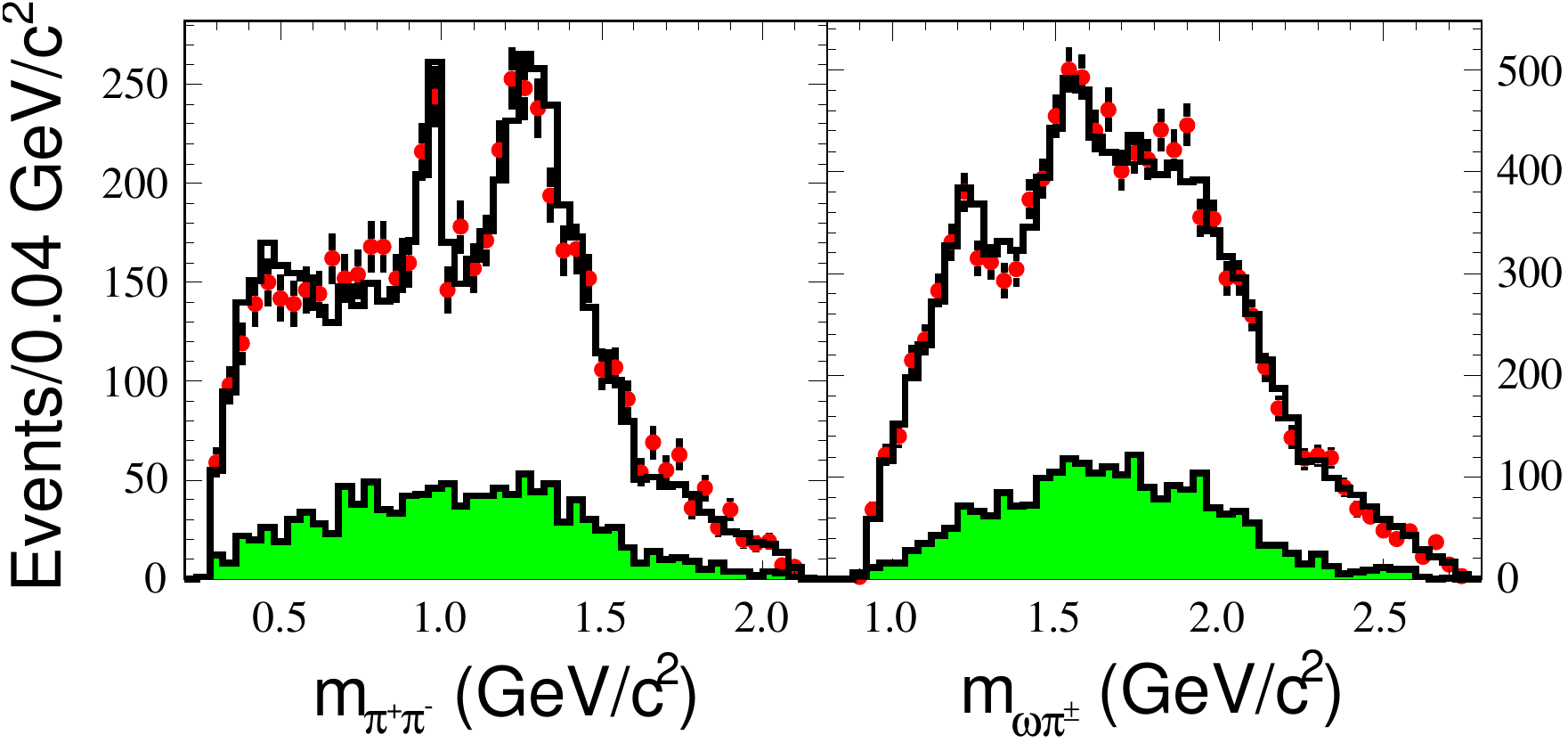}
}
\caption{Projections of the fit results for mass spectra $m_{\pp}$ and $m_{\omega\pi^\pm}$ for group B data sets. Dots with error bars are the data, histograms are the total fit results, and shaded histograms (green) are the background estimated by the $\omega$ mass sideband.}
\label{fig:ToTB}
\end{center}
\end{figure}

Signal yields for the data sets are calculated using Eq.~(\ref{yieldsFormula}), and their statistical uncertainties are derived using Eq.~(\ref{staterr}), which includes the correlation among parameters. The signal yields are given in Table~\ref{tabsigyields}. Projections of invariant masses for individual energy points are shown in Fig.~\ref{fig:A}. The combined energy points of mass projections of fit results for groups A and B are shown in Fig.~\ref{fig:ToTA} and Fig.~\ref{fig:ToTB}, respectively.


\section{Born cross section}
\subsection{ISR correction factor}
In the direct $\ee$ collision experiments, the observed cross section, $\sigma_{\text{obs}}(s)$, at the c.m. energy $\sqrt s $ for \mbox{$\ee\to\omega\pp$} is the Born cross section, $\sigma_0(s)$, convolved with the ISR function $W(s,x)$~\cite{exlusgen1,exlusgen2}. To unfold the Born cross section, the ISR correction factor should be defined as
\begin{eqnarray}
1+\delta &=& {\sigma_{\text{obs}}(s)\over \sigma_0(s)},
\end{eqnarray}
with
\begin{eqnarray}
\sigma_{\text{obs}}(s)&=& \int_{{M_{\text{th}}}}^{\sqrt{s}}W(s,x){\sigma_0[s(1-x)]\over |1-\Pi(\sqrt{s})|^2}dx,
\end{eqnarray}
where $\Pi(\sqrt{s})$ is the vacuum polarization function. We use the calculated results, including the leptonic and hadronic parts both in space- and time-like regions \cite{vp,vp0,vp1,vp2,vp3}. The $M_{\text{th}}$ corresponds to the $\omega\pp$ mass threshold, and $x$ is the ratio of the ISR photon energy to the beam energy.

Calculations of the ISR correction factor and the MC event generation are consistently performed using the generator model ``ConExc''~\cite{exlusgen2}, and the
dressed cross sections $\sigma_0^{D}(s)$ are taken from the BaBar experiment ($\sqrt s<2.000$ GeV)~\cite{BABAR1} and this measurement~\mbox{($\sqrt s>2.000$ GeV)}. To achieve stable cross sections in this measurement, the procedure of the ISR correction factor calculation and the MC event generation for the Born cross section calculation are iterated serval times. The iteration is stopped if the updated Born cross section reaches the accuracy within statistical uncertainty. The ISR correction factors for various energy points are given in Table~\ref{CrsCalResults}.

\begin{figure*}
\subfigure{
		\includegraphics[width=8cm,height=8.cm]{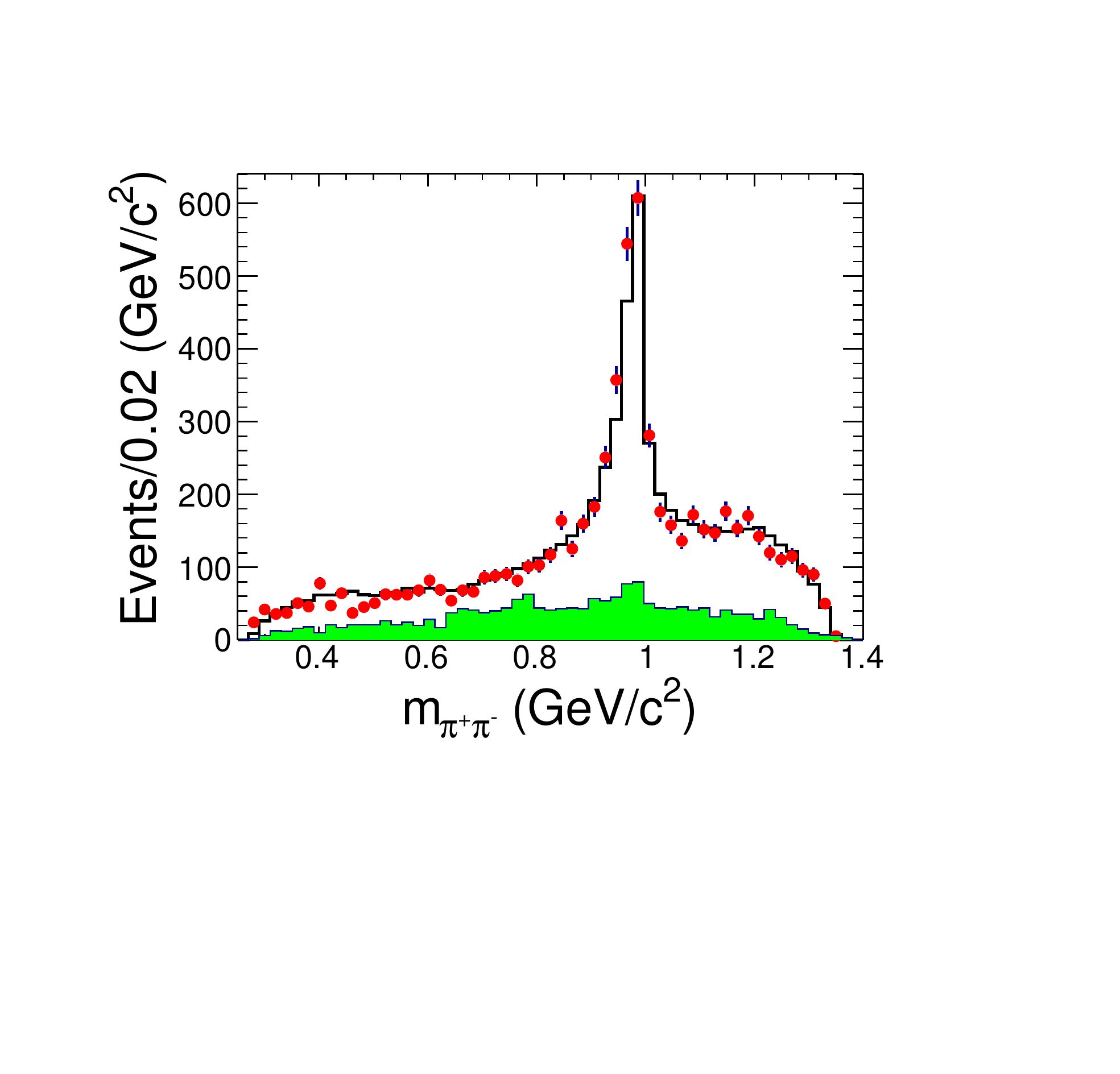}
\hspace{-1.5cm}
		\includegraphics[width=8cm,height=8.cm]{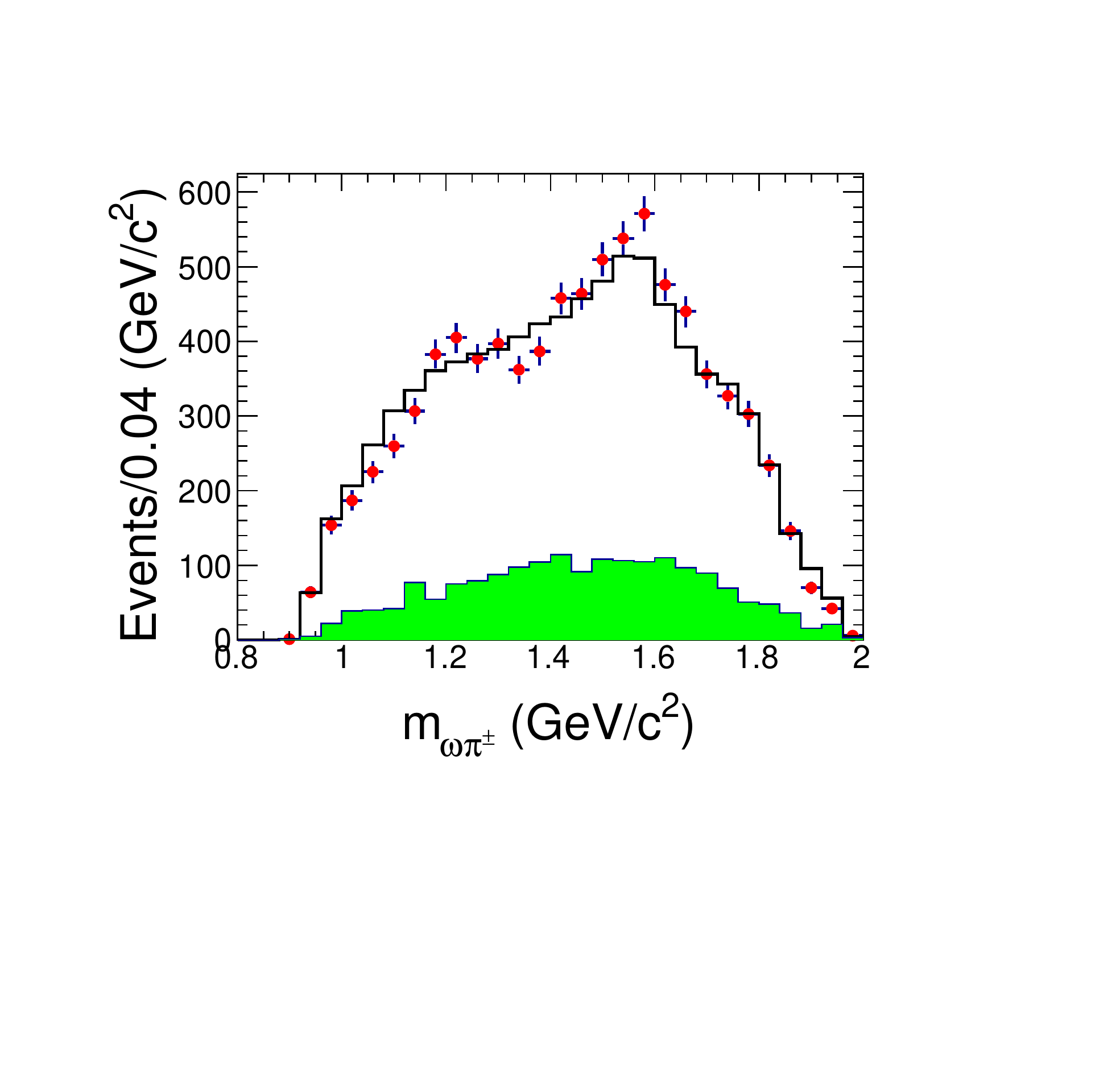}
}
\vspace{-2.5cm}
\caption{\label{mcevents} Comparison of $m_{\pi^{+}\pi^{-}}$ and $m_{\omega\pi^{\pm}}$ mass spectra between data and MC simulation at $\sqrt{s}=$~2.1250 GeV. Dots with error bars are the data, histograms are the MC simulation, and shaded histograms (green) are the background estimated by the $\omega$ mass sideband.}
\end{figure*}

In the MC event generation, the ISR photon is characterized by the soft energy and beam collinear distribution.~Its angular dependence is sampled with the Bonneau and Martin formula with an accuracy of up to the $m_e^2/s$ term;
\begin{eqnarray}
P(\theta,x)&=&\frac
{\sin^2{\theta}-\frac{x^2\sin^4{\theta}}{2(x^2-2x+2)}-
\frac{m_\mathrm{\it e}^2}{E^2}~\frac{(1-2x)\sin^2{\theta}-x^2\cos^4{\theta}}{x^2-2x+2}}
{\left ( \sin^2{\theta}+\frac{m_\mathrm{e}^2}{E^2}\cos^2{\theta}
\right )^2 },
\end{eqnarray}
where $\theta$ is the polar angle of the ISR photon, $m_e$ is the electron mass, and {\it E} is the beam energy. After the photon emission, the $\gamma^*\to\omega\pp$ events are generated with the amplitude model with parameters fixed to the fitted values so that the measured intermediate states are inclusively simulated. Figure~\ref{mcevents} shows an example of the MC simulation at \mbox{$\sqrt s=2.125$ GeV}, in which we can observe a good agreement between data and MC simulation in the mass distribution.

\subsection{Cross section for $\ee\to\omega\pp$}\label{CrsOfWpp}
\begin{figure}[htbp]
\begin{center}
\subfigure{
\includegraphics[scale=0.5]{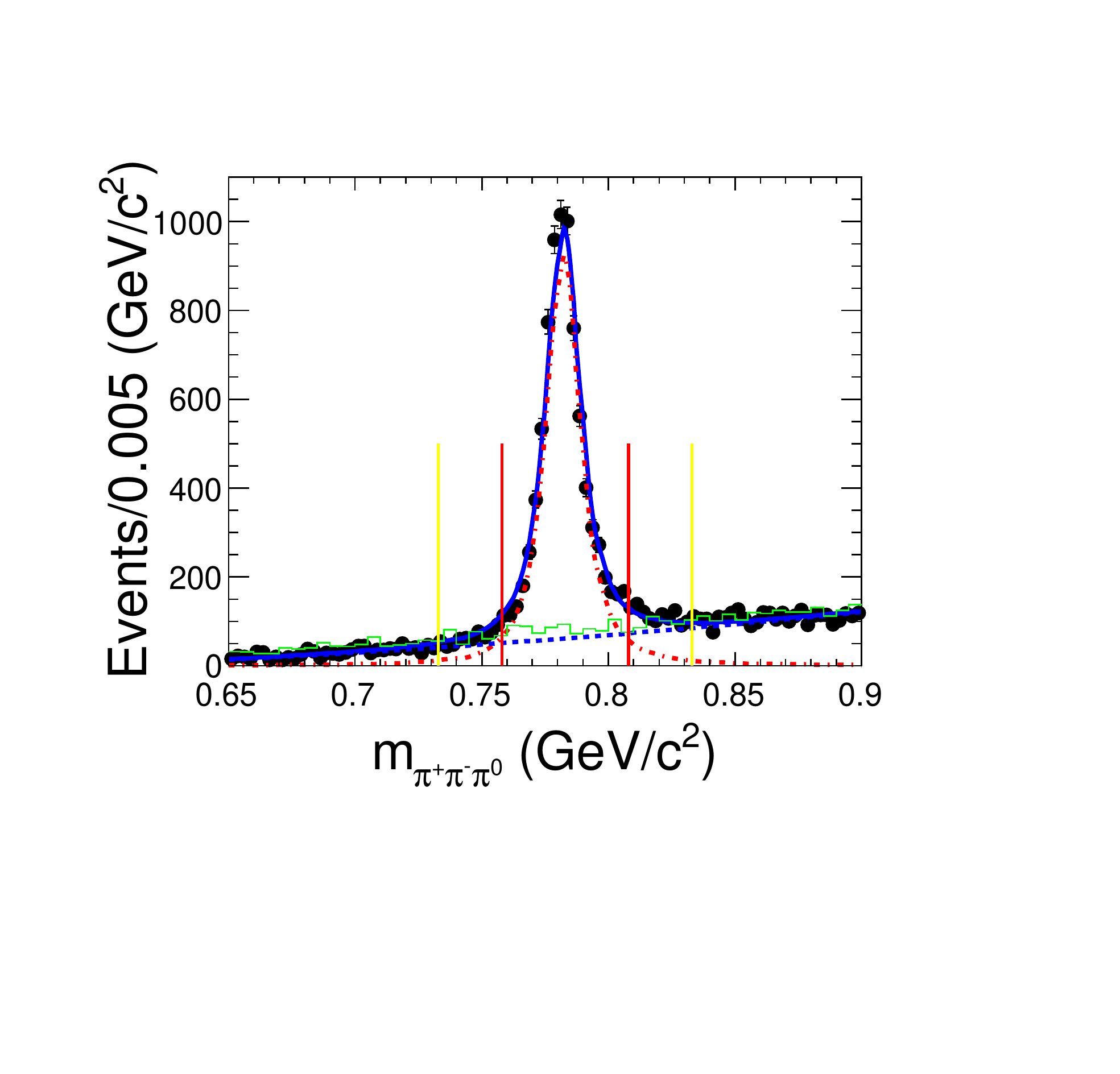}
}
\vspace{-2.5cm}
\caption{\label{fig:wfit} The $\pi^{+}\pi^{-}\pi^{0}$ invariant mass spectrum of the process $\ee\to \omega\pi^{+}\pi^{-}$ at \mbox{$\sqrt{s}=$~2.1250 GeV}. The dots with error bars are the data, the red dotted curve is the signal, and the blue dashed line is the background. The blue solid curve is the best fit result. The green histogram is from the MC simulation of the non-resonant contribution from $\ee \to \pi^{+}\pi^{-}\pi^{+}\pi^{-}\pi^{0}$. The range between the two red vertical solid lines is regarded as the signal region, and those between the yellow and red vertical lines on each side of the signal peak are regarded as the sideband regions.}
\end{center}
\end{figure}

The signal yields are determined by an unbinned maximum likelihood fit to the $\pp\pi^{0}$ invariant mass distribution of the $\omega$ candidates. In the fit, the signal shape is modeled with the MC histogram smeared with a Gaussian function for the resolution difference between data and MC simulation. The background is dominated by the process $e^{+}e^{-}\rightarrow2(\pi^{+}\pi^{-})\pi^{0}$, and the corresponding shape is described by a second-order Chebychev polynomial. The parameters of the Gaussian function, the background polynomial and the yields of signal and background are floated. Figure~\ref{fig:wfit} shows the fit results at $\sqrt {s} =$ 2.125 GeV. The Born cross section is calculated from
\begin{eqnarray}
\sigma^B=\frac{N_{\rm sig}-N_{\rm mis}}{\mathcal{L}\cdot \epsilon \cdot Br(\omega\rightarrow\pi^{+}\pi^{-}\pi^{0}) \cdot Br(\pi^{0}\rightarrow\gamma\gamma)\cdot(1+\delta)},
	\label{CrsCal}
\end{eqnarray}
where $N_{\rm sig}$ is the number of signal events obtained by fitting the $\pi^{+}\pi^{-}\pi^{0}$ mass distribution, $N_{\rm mis}$ the mis-combination yield, $\mathcal{L}$ the integrated luminosity, and $\epsilon$ the detection efficiency obtained from the MC simulation according to the PWA result. ${\it Br}(\omega\rightarrow\pi^{+}\pi^{-}\pi^{0}$)  and $\it Br(\pi^{0}\rightarrow\gamma\gamma$) are the branching fractions quoted from the PDG~\cite{PDG2020}, and $(1+\delta)$ represents the correction factors due to the ISR effect and vacuum polarization.
The resulting Born cross sections and related variables are summarized in Table~\ref{CrsCalResults}.

\begin{table*}
\begin{center}
\caption{Summary of the signal yield~($N_{\rm sig}$), number of mis-combination events~($N_{\rm mis}$), integrated luminosity $\mathcal{L}$, ISR correction factor~($1+\delta$), detection efficiency~($\epsilon$), and total Born cross section of~$\ee\to\omega\pi^{+}\pi^{-}$~($\sigma^{B}$) at various energy points~$\sqrt{s}$. The uncertainties of $N_{\rm sig}$ and $N_{\rm mis}$ are only statistical. The first and second uncertainties of Born cross sections are statistical and systematic, respectively.}
\vspace{0.3cm}
\begin{small}
\begin{tabular}{C{1.2cm}C{0.8cm}PR{0.4cm}C{0.6cm}PR{0.2cm}R{0.6cm}@{.}L{0.8cm}C{1.0cm}C{1.0cm}R{0.8cm}PR{0.6cm}PR{0.4cm}r}
                \hline\hline
$\sqrt{s}~$(GeV) &\multicolumn{2}{c} {N$_{\rm sig}$} &\multicolumn{2}{c} {N$_{\rm mis}$}&\multicolumn{2}{c}{$\mathcal{L}$(pb$^{-1}$)}&$1+\delta$ & $\epsilon$ & \multicolumn{4}{c}{$\sigma^{B}$(pb)} \\ \hline
2.0000&1008& 37& 10&3&  10&1&  1.1819&0.1775& 534.3&19.8&29.9&\\
2.0500&239 & 18& 3 &2&  3&4&   1.1912& 0.1615& 421.3&32.0&23.4&\\
2.1000&952 & 37& 9 &3&  12&2&  1.1707& 0.1764& 424.6&16.7&22.7&\\
2.1250&8013& 103&80&9& 109&0&  1.1715&0.1704& 413.6&5.4&22.3 &\\
2.1500&207 & 17& 2 &1&  2&9&   1.1649& 0.1587& 446.4&37.4&29.1&\\
2.1750&863 & 34& 9 &3&  10&6&  1.1789&0.1646& 471.0&18.8&24.6&\\
2.2000&971 & 36& 10&3&  13&7&  1.1858&0.1631& 411.5&15.4&21.4&\\
2.2324&821 & 32& 8 &3&  11&9&  1.1986&0.1556& 415.6&16.4&23.0&\\
2.3094&856 & 35& 9 &3&  21&1&  1.2543&0.1503& 241.6&10.0&14.4&\\
2.3864&664 & 30& 7 &3&  22&6&  1.3953&0.1470& 160.8&7.3&~8.4 &\\
2.3960&1943& 53&19 &4&  66&9&  1.4114&0.1324& 174.9&4.8&~9.5 &\\
2.6444&624 & 29& 6 &2&  33&7&  1.4456&0.1228& 117.2&5.5&~6.3 &\\
2.6464&606 & 29& 6 &2&  34&1&  1.4463&0.1481&  93.2&4.5&~5.3 &\\
2.9000&999 & 36&10 &3&  106&0& 1.6241&0.1035&  63.0&2.3&~3.6 &\\
2.9500&136 & 14& 2 &1&  16&0&  1.5524&0.1044&  59.9&5.9&~3.9 &\\
2.9810&130 & 13& 2 &1&  16&1&  1.5029&0.1101&  55.3&5.6&~3.3 &\\
3.0000&119 & 12& 1 &1&  15&9&  1.5038&0.1076&  52.7&5.4&~3.5 &\\
3.0200&134 & 13& 2 &1&  17&3&  1.5308&0.1058&  54.1&5.3&~3.2 &\\
3.0800&543 & 28& 5 &3& 126&2&  1.7893&0.0776&  35.2&1.8&~2.6 &\\
 \hline \hline
\end{tabular}
\label{CrsCalResults}
\end{small}
\end{center}
\end{table*}

\subsection{Cross sections for the intermediate subprocesses}
The Born cross sections for the intermediate subprocesses are determined with the total Born cross sections of the~$\ee\to\omega\pi^{+}\pi^{-}$ process and the PWA results. Using a signal MC sample of events generated uniformly in the phase space without the ISR effect, the ratio of the Born cross section for a subprocess $R$ to the total cross section is determined.
\begin{equation}
r_R = {\sigma_R\over \sigma_{\text{tot}}},
\end{equation}
where $\sigma_R$ and $\sigma_{\text{tot}}$ are obtained by the partial wave amplitude weighted MC sample. The Born cross sections of the intermediate subprocess at each of the selected energy points are given in Table~\ref{ResCrs}.

\begin{table*}
\caption{\label{ResCrs} Born cross sections for the subprocesses at various selected energy points in unit of pb, where the first and second uncertainties are statistical and systematic, respectively. }
\vspace{0.3cm}
\begin{tabular}{C{1.0cm}R{.8cm}PR{0.6cm}PR{0.5cm}R{1.0cm}PR{0.7cm}PR{0.7cm}R{1.2cm}PR{0.7cm}PR{0.7cm}r} \hline\hline
$\sqrt{s}~$(GeV) &\multicolumn{2}{c} {$\omega f_{0}(500)$} &\multicolumn{3}{c} {$\omega f_{0}(980)$} &\multicolumn{3}{c}{$\omega f_{0}(1370)$} \\ \hline
2.0000&  37.7&15.2&5.0& 237.3&22.6&31.3& 114.0&28.6&15.1 &\\
2.1000&  34.7&15.3&5.5&  117.1&9.5&18.6& 60.3&14.1&9.6 &\\
2.1250&  45.6& 4.9&5.0& 144.3&6.6&15.7& 49.8&7.5&5.4 &\\
2.1750&  91.1&16.6&6.7& 210.1&25.7&38.5& 90.3&37.5&16.5 &\\
2.2000&  78.2&16.4&7.0& 133.4&15.8&29.0& 155.1&23.9&33.7& \\
2.2324&  63.8&10.0&2.5& 154.1&16.3&30.1& 119.2&21.5&23.3& \\
2.3094& 73.1&14.9&3.0& 52.4&20.6&9.3& 42.4&35.5&7.5 & \\
2.3864&  34.5&6.4&5.8& 30.7&13.2&5.2& 21.8&9.2&3.7 &\\
2.3960&  50.4&4.8&1.5& 48.1&16.5&11.0& 45.6&7.1&10.5& \\
2.6444&  23.1&4.8&2.2& 10.1&3.5&1.0& 11.6&5.3&1.1 &\\
2.6464&  20.3&3.1&3.8& 16.7&4.2&3.2& 19.6&5.0&3.7 &\\
2.9000& 15.7&2.0&1.7&  2.8 &1.4&0.3& 3.8&2.0 &0.4 &\\ \hline
\end{tabular}
\begin{tabular}{C{1.0cm}R{1.cm}PR{0.7cm}PR{0.6cm}R{1.0cm}PR{0.6cm}PR{0.5cm}R{1.0cm}PR{0.6cm}PR{0.5cm}r}
$\sqrt{s}~$(GeV) &\multicolumn{2}{c} {$\omega f_{2}(1270)$} &\multicolumn{4}{c}{$b_{1}^{\pm}(1235)\pi^{\mp}$} &\multicolumn{4}{c}{non-resonant $\omega \pi^{+}\pi^{-}$} \\ \hline
2.0000&   107.6&17.3&14.2& 17.9&5.7&2.4 &1.9&1.9&0.1 &\\
2.1000&    71.3&13.8&11.3& 10.2& 7.3&1.6&6.0&5.6&0.3 & \\
2.1250&    48.0&5.4&5.3& 19.9&2.6&2.2 &1.6&0.3&0.1 &\\
2.1750&    56.1&19.9&10.3& 39.1&8.1&7.2&3.6&1.9&0.2 & \\
2.2000&   90.3&18.8&19.6& 26.8&15.7&5.8 &6.1&4.8&0.3 &\\
2.2324&    49.2&14.2&9.6& 15.2&5.7&3.0&8.3&1.2&0.5 & \\
2.3094&  43.4&10.1&7.7& 6.3&3.6&1.1 &0.8&0.4&0.1 & \\
2.3864&   22.4&4.4&3.8& 7.9&2.7&1.3 &9.6&1.6&0.5 &\\
2.3960&   13.2&3.9&3.0& 15.2&3.1&3.5&0.7&0.5&0.1 & \\
2.6444&   15.5&4.7&1.5& 10.2& 2.7&1.0&2.4&0.5&0.1 & \\
2.6464&   13.5&2.8&2.6& 14.8&2.0&2.8 &1.9&1.3&0.1 &\\
2.9000&  4.7&1.6&0.5& 9.0&1.1&1.0 &2.9&0.6&0.2 &\\
\hline\hline
\end{tabular}
\end{table*}

\section{\label{sec:level6}Systematic uncertainties}
\subsection{Uncertainties for intermediate subprocesses}

The uncertainties for the measurements of intermediate subprocesses due to the PWA originate from the parameterizations of the $f_0(500)$ and $f_0(980)$ states,  the resonance parameters, insignificant contributions for some intermediate states, and background contamination.
\begin{itemize}
\item {In the nominal fit, we take the $f_0(500)$ line shape as that used by the E791 Collaboration. The uncertainty associated
with this parametrization is estimated by replacing it with those in Ref.~\cite{besiib}. The difference in the signal yield
of each intermediate state is considered as the systematic uncertainty.}
\item {The parameters of the $f_0(980)$ Flatt\^e~formula in the nominal fit are fixed to the measurements \cite{besiia,besiib}.
The uncertainties are estimated by varying the resonance parameters by one standard deviation. The differences in the signal yields are considered as the systematic uncertainties.}
\item {The masses and widths of intermediate states $f_0(1370)$, $f_2(1270)$, and $b_1(1235)$ are fixed to the PDG values~\cite{PDG2020} in the nominal fit. The associated uncertainties are estimated by varying masses and widths by $\pm 1\sigma$ respectively. The differences are taken as the corresponding systematic uncertainties.}
\item {The two resonances $\rho(1450)^\pm$ and $\rho(1570)^\pm$ are insignificant and not included in the baseline solution. The differences of the fits with and without the two resonances in the signal yields are assigned as systematic uncertainties.}
\item {To estimate the uncertainties associated with background contamination, alternative fits are performed
by changing the number of background events with one standard deviation for various energy points. The differences in the signal yields are considered as the systematic uncertainties.}
\end{itemize}
All these uncertainties are added in quadrature to provide the total systematic uncertainties of intermediate states (Table~\ref{ResCrs}).

\begin{table*}
\caption{\label{tab:table3} Systematic uncertainties~(in unit of $\%$) of the $e^{+}e^{-}$$\rightarrow$ $\omega \pi^{+}\pi^{-}$ cross section measurements at various energy points ~($\sqrt{s}$). The main sources are from track efficiency~(Trk), photon detection efficiency~($\epsilon_{\gamma}$), 4C kinematic fit~(4C), $\pi^{0}$ mass window~($\pi^{0}$), ISR effect, signal shape (SS) and background shape (BS), fit range (Fit), luminosity measurement ($\mathcal{L}$), $\omega$ decay branching fraction~($Br$), MC statistics~($\Delta_{\rm MC}$), and MC model. Meanwhile, the total uncertainty for each point~(Total) is obtained by summing the individual contributions in quadrature.}
\vspace{0.3cm}
\begin{tabular}{cccccccccccccc}
\hline\hline
$\sqrt s$~(GeV)& Trk  & $\epsilon_{\gamma}$ &4C  &$\pi^{0}$& ISR  & SS & BS & Fit & $\mathcal{L}$ &   $Br$ & $\Delta_{\rm MC}$& Model &Total \\ \hline
2.0000& 4.0& 2.0&   1.1& 1.2&1.0&1.9&  0.1& 0.2&    1.0& 0.8& 0.2& 1.5 & 5.6 \\
2.0500& 4.0& 2.0&   1.1& 1.0&1.0&2.0&  0.6& 0.3&    1.0& 0.8& 0.2& 1.3 & 5.6 \\
2.1000& 4.0& 2.0&   1.2& 1.2&1.0&1.4&  0.1& 0.5&    1.0& 0.8& 0.2& 0.7& 5.3 \\
2.1250& 4.0& 2.0&   1.1& 1.1&1.0&1.7&  0.1& 0.3&    1.0& 0.8& 0.2& 0.9& 5.4 \\
2.1500& 4.0& 2.0&   1.1& 2.0&1.0&3.5&  0.2& 0.5&    1.0& 0.8& 0.2& 1.2& 6.5 \\
2.1750& 4.0& 2.0&   1.1& 0.3&1.0&1.4&  0.3& 0.7&    1.0& 0.8& 0.2& 0.6& 5.2 \\
2.2000& 4.0& 2.0&   1.0& 1.1&1.0&1.0&  0.4& 0.1&    1.0& 0.8& 0.2& 1.0& 5.2 \\
2.2324& 4.0& 2.0&   1.1& 0.9&1.0&0.9&  0.9& 1.9&    1.0& 0.8& 0.2& 0.4& 5.5 \\
2.3094& 4.0& 2.0&   1.3& 1.0&1.0&1.7&  0.3& 2.4&    1.0& 0.8& 0.2& 0.8& 5.9 \\
2.3864& 4.0& 2.0&   1.4& 0.7&1.0&0.8&  0.1& 0.2&    1.0& 0.8& 0.3& 1.0& 5.2 \\
2.3960& 4.0& 2.0&   1.2& 0.2&1.0&1.8&  0.9& 0.9&    1.0& 0.8& 0.3& 0.3& 5.4 \\
2.6444& 4.0& 2.0&   1.1& 0.4&1.0&1.7&  0.7& 0.9&    1.0& 0.8& 0.3& 0.5& 5.3 \\
2.6464& 4.0& 2.0&   1.3& 0.4&1.0&1.8&  1.1& 1.1&    1.0& 0.8& 0.3& 1.3& 5.6 \\
2.9000& 4.0& 2.0&   1.6& 2.0&1.0&1.5&  0.5& 0.2&    1.0& 0.8& 0.3& 0.5& 5.7 \\
2.9500& 4.0& 2.0&   1.5& 2.0&1.0&3.5&  0.4& 0.6&    1.0& 0.8& 0.3& 0.5& 6.5 \\
2.9810& 4.0& 2.0&   1.5& 2.0&1.0&2.0&  0.3& 0.2&    1.0& 0.8& 0.3& 0.8& 5.9 \\
3.0000& 4.0& 2.0&   1.4& 3.0&1.0&3.0&  0.6& 0.7&    1.0& 0.8& 0.3& 0.7& 6.7 \\
3.0200& 4.0& 2.0&   1.5& 1.0&1.0&2.5&  0.5& 0.2&    1.0& 0.8& 0.3& 1.0& 5.8 \\
3.0800& 4.0& 2.0&   1.3& 2.0&1.0&4.5&  0.8& 0.6&    1.0& 0.8& 0.4& 0.9& 7.2 \\
\hline\hline
\end{tabular}
\end{table*}
\subsection{Uncertainties for the measurements of $\sigma(\ee\to\omega\pp)$}
The sources of systematic uncertainties associated with the Born cross section measurements include the tracking efficiency, photon detection, kinematic fit, $\pi^{0}$ mass window requirement, fit procedure, radiative correction, luminosity measurement, and branching fraction of $\omega\rightarrow\pi^{+}\pi^{-}\pi^{0}$. These sources are described as follows:
\begin {itemize}
\item {The uncertainty due to the tracking efficiency is estimated as $1.0\%$ per track~\cite{TrackErr}. }
\item {The uncertainty due to the photon detection is $1.0\%$ per photon~\cite{PhotoErr}. }
\item {The uncertainty associated with the kinematic fit originates from the inconsistency between the data and MC simulation of the track helix parameters. Following the procedure described in Ref.~\cite{4CfitErr}, half of the difference in the detection efficiencies with and without the helix parameter correction is regarded the systematic uncertainty.}
\item {The $\pi^0$ candidate is selected by requiring $|M_{\gamma\gamma}-0.135|$<0.015 GeV$/c^{2}$. By changing this requirement to $|M_{\gamma\gamma}-0.135|$<0.018 GeV/$c^{2}$ or $|M_{\gamma\gamma}-0.135|$<0.013 GeV$/c^{2}$, the larger difference in the measured cross section is considered as systematic uncertainty.}
\item {The uncertainties due to the choices of signal shape, background shape, and fit range are estimated by varying the signal function from the Breit-Wigner to the MC simulated shape convolved with a Gaussian resolution function. Another technique is varying the background function from the second-order to third-order Chebychev polynomial and by extending the fit range from (0.65, 0.89) GeV$/c^{2}$ to \mbox{(0.64, 0.90) GeV$/c^{2}$}. The differences in the signal yields are considered as systematic uncertainty.}
\item {The ISR correction factor is calculated with the line shape of Born cross sections, which is smoothed by fitting the cross section distribution with five Gaussian functions. The signal MC samples with the ISR effect are generated according to the input line shape. Using the updated MC samples, we re-estimate the efficiency and ISR correction factor and then update the cross section. Iterations are performed until the results are stable. Finally, the difference between the last two iterations, $1.0\%$, is considered the corresponding systematic uncertainty for all the energy points.}
\item {The integrated luminosities of these data sets are measured using large angle Bhabha scattering events, with a uncertainty about $1.0\%$~\cite{LumErr} is propagated to the cross section measurements. The uncertainty in $Br(\omega\rightarrow\pi^{+}\pi^{-}\pi^{0}$) is quoted as $0.8\%$ from the PDG~\cite{PDG2020}.}
\item {The uncertainty from the MC statistics is estimated by the number of generated events, calculated by $\Delta_{\rm MC}$=$\sqrt{(1-\epsilon)/\epsilon}/\sqrt{N}$, where $N=$1000000 is the number of generated events at each energy points.}
\item{The uncertainty due to the PWA MC model is investigated by smearing masses and widths of intermediate states per event within $\pm 1\sigma$, assuming that they follow Gaussian distribution. The difference in the detection efficiencies is considered the systematic uncertainty, which is about $1\%$. The effect due to the insignificant intermediate states, such as $\rho(1450)$ and $\rho(1570)$, is investigated by adding them to the nominal fit. Moreover, the difference in the detection efficiencies with and without these two intermediate states is regarded the systematic uncertainty, which varies from $0.2\%$ to $1\%$ depending on the energy point. The total uncertainty due to the MC model is obtained by adding these uncertainties in quadrature.}
\end{itemize}
The systematic uncertainties for all energy points are listed in Table~\ref{tab:table3}.
These uncertainties are assumed independent and summed in quadrature.

\section{Fit to the line shapes}
\subsection{\label{sec:leve71}Line shape of $e^{+}e^{-}\rightarrow\omega\pi^{+}\pi^{-}$}
\begin{figure}
\begin{center}
\subfigure{
\hspace{-1.0cm}
\includegraphics[scale=0.45]{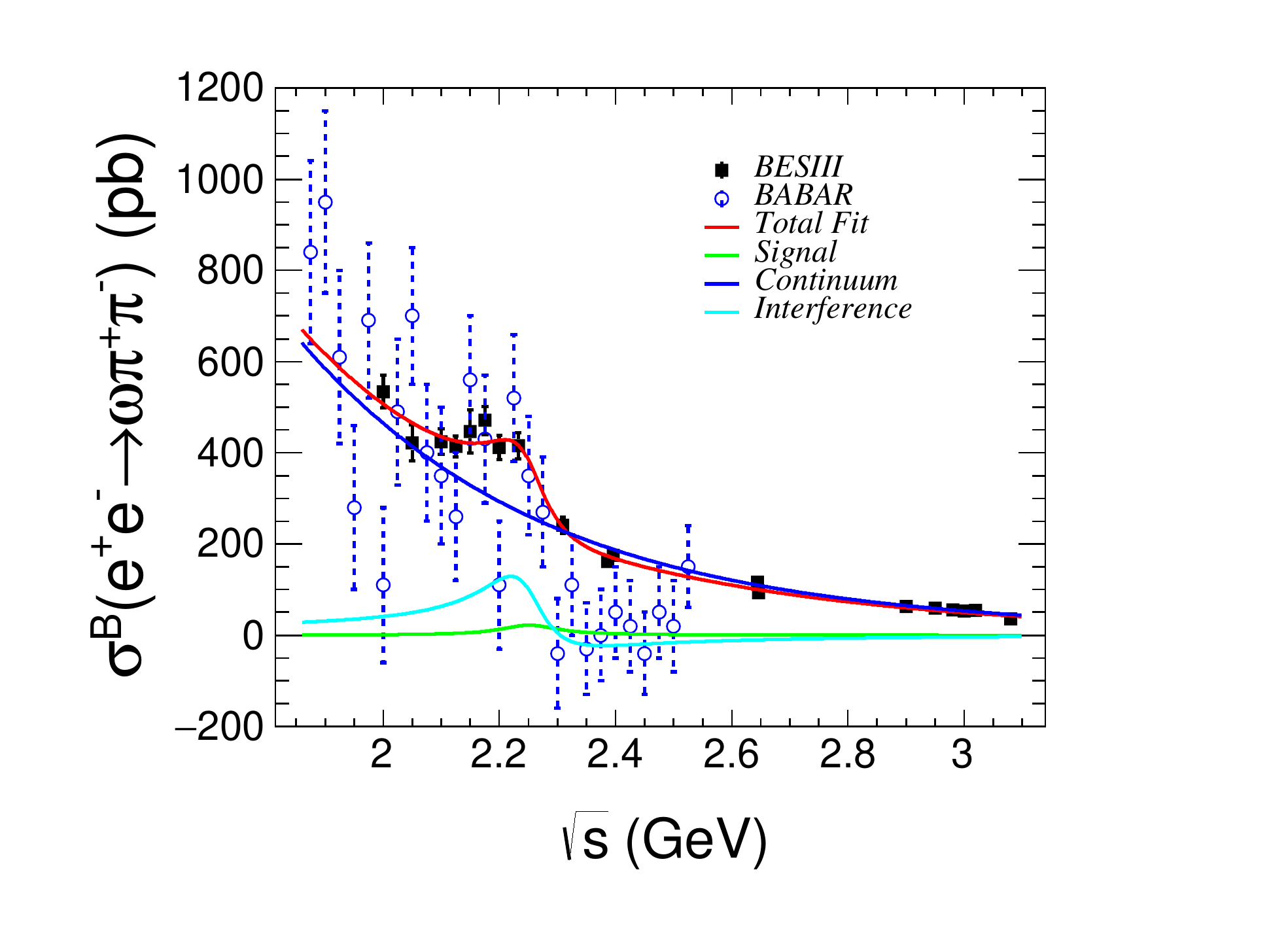}
\hspace{-1.2cm}
\includegraphics[scale=0.45]{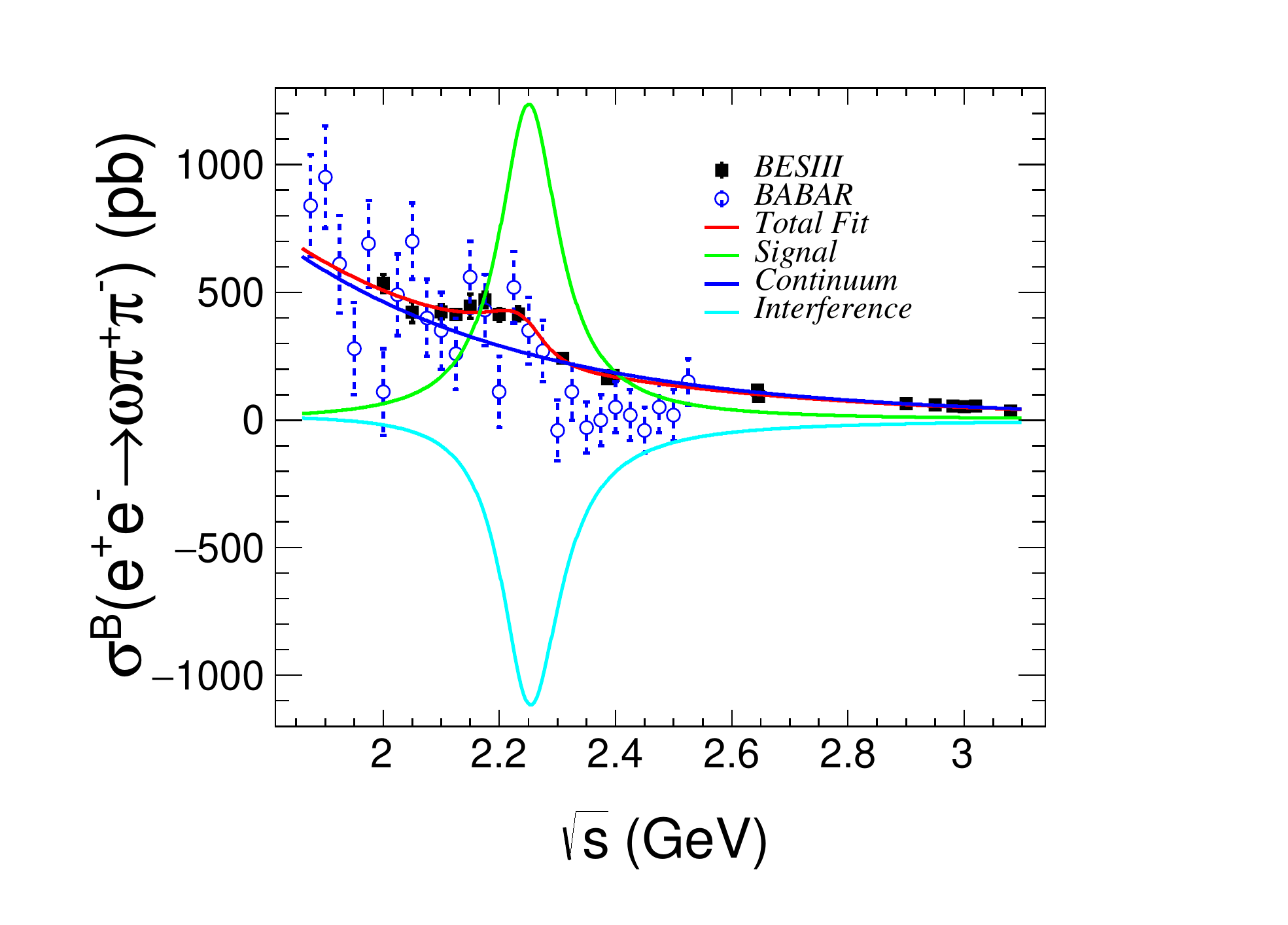}
}
\caption{\label{fig:CSfit} Fits to the measured $e^{+}e^{-}\rightarrow\omega\pi^{+}\pi^{-}$ cross sections. The dots with error bars are the results obtained in the work, where the error bars include both statistical and systematic uncertainties. Circles with error bars (hatched line) are the results of the BaBar experiment. The red solid curve is the total fit result. The blue solid curve represents the continuum component, the green curve stands for the resonant component, and the cyan line is the interference of the two components. The two plots are the results of Solutions I (left) and II (right), respectively.}
\end{center}
\end{figure}

The measured Born cross sections are shown in Fig.~\ref{fig:CSfit}, where a clear structure is observed around 2.25 GeV.
A minimized $\chi^{2}$ fit incorporating the correlated and uncorrelated uncertainties is performed for the measured cross section with the following function:
\begin{eqnarray}
\sigma^B(\sqrt{s})&=|&\sigma_\textrm{r}(\sqrt{s})e^{i\phi} + \sigma_\textrm{c}(\sqrt{s})|^{2},
\end{eqnarray}
where $\phi$ is the phase of the resonance structure relative to the continuum components and $\sigma_\textrm{r}(\sqrt{s})$ and $\sigma_\textrm{c}(\sqrt{s})$ are the cross sections for the resonance and continuum components, respectively. The resonant component $\sigma_\textrm{r}(\sqrt{s})$ is parameterized as
\begin{eqnarray}\label{SimuFitEq}
\sigma_\textrm{r}(\sqrt{s})&=&\frac{m_{r}}{\sqrt{s}} \frac{\sqrt{12\pi C\Gamma^{ee}_{r}Br\Gamma_{r}}}{s-m^{2}_{r}+im_{r}\Gamma_{r}}\sqrt{\frac{P(\sqrt{s})}{P(m_{r})}},
\end{eqnarray}
where $m_{r}$ and $\Gamma_{r}$ are the mass and width of the resonant structure near 2.25 GeV, respectively. Parameter $\Gamma^{ee}_{r}\cdot Br$ is the product of the partial width of the resonance decaying to the $e^{+}e^{-}$ and the branching fraction to the $\omega\pi^{+}\pi^{-}$ final state. $m_{r}$, $\Gamma_{r}$ and $\Gamma^{ee}_{r}\cdot Br$ are free parameters in the fit. $C$ is a conversion constant that equals to \mbox{$3.893\times 10^{5}$ nb{$\cdot$}GeV$^{2}$}~\cite{BABAR1}. The continuum component is parameterized as
\begin{eqnarray}\label{SimuFitEq1}
\sigma_\textrm{c}(\sqrt{s})&=& \frac{a\sqrt{P(\sqrt{s})}}{\sqrt{s}^{b}},
\end{eqnarray}
 where a and b are free parameters and $P(\sqrt{s})$ is the phase factor. For the two-body process, $P(\sqrt s)$ is taken as $p/\sqrt s$, where $p$ is the magnitude of the momentum for one of the two particles. For the three-body final state of $\omega\pi^+\pi^-$,
\begin{equation}
P(\sqrt{s})=\frac{1}{4s}\int_{2m_{\pi}}^{\sqrt s-m_{\omega}}\sqrt{(m_{\pp}^2-4m_{\pi}^2)[s-(m_{\pp}+m_{\omega})^2][s-(m_{\pp}-m_{\omega})^2]}dm_{\pp},
\end{equation}
where $m_{\omega}$ and $m_{\pp}$ are the mass of $\omega$ particle and the invariant mass of $\pp$, respectively.
The results of the fits to the measured cross section are shown in Fig.~\ref{fig:CSfit}. The data points of the BaBar experiment are plotted to overlap for comparison. ~The fit has two solutions with equally good fit quality and identical mass and width of the resonance, but with different phases and $\Gamma^{ee}_{r}\cdot Br$. We observe a structure around 2.25 GeV, denoted as $X(2230)$. The statistical significance of $X(2230)$ is 10.3$\sigma$, which is determined from the change in the $\chi^2$ value with and without it in the fit versus the change of the number of degrees of freedom.
The best fit provides a fit quality of $\chi^{2}$/n.d.f = 20.6/13. The fit parameters of the two solutions are listed in Table~\ref{tab:table4}.~Due to the interference effect, the parameter $b$ is fitted as $b=4.4\pm0.1$ in both two solutions. Note that this is different from the single virtual photon contribution where $b=1$.
\begin{table*}[htbp]
\begin{center}
\caption{\label{tab:table4} Resonance parameters obtained in the fit to the $\ee\to \omega \pi^{+}\pi^{-}$ cross section.~The uncertainties are statistical only.}
\vspace{0.3cm}
\begin{tabular}{L{2.3cm}R{1.0cm}PR{0.7cm}R{1.3cm}PR{0.7cm}r} \hline\hline
Parameter  &\multicolumn{2}{r} {Solution I} &\multicolumn{2}{r}{Solution II}\\\hline
$m_{r}$ (MeV/$c^{2}$)&\multicolumn{4}{c}{2250~$\pm$~25}& \\
$\Gamma_{r}$ (MeV) &\multicolumn{4}{c}{125~$\pm$~23}&\\
$\Gamma^{ee}_{r}\cdot Br$(eV)&0.9&0.4&52.9 &17.0&\\
$\phi$ (rad.) &2.4&0.3&$-$1.8 & 0.1&\\
$a(10^{3})$(pb$^{1/2}$)  &\multicolumn{4}{c}{1.1~$\pm$~0.2}&\\
$b$  &\multicolumn{4}{c}{4.4~$\pm$~0.1}& \\
Significance & \multicolumn{4}{c}{10.3$\sigma$} \\ \hline\hline
\end{tabular}
\end{center}
\end{table*}

To estimate the systematic uncertainties for the resonant parameters, an alternative fit is performed by parameterizing the continuum component with an exponential function~\cite{Bes3PRL} to determine the uncertainties associated with the continuum component:\\
\begin{eqnarray}\label{continuum}
\sigma_{c}(\sqrt{s})=\sqrt{P(\sqrt{s})e^{p_{0}u}p_1},
\end{eqnarray}
where $p_{0}$~and~$p_{1}$ are float parameters and $u=\sqrt{s}-(2m_{\pi}+m_{\omega})$, which the masses of $m_{\pi}$ and $m_{\omega}$ take from the PDG~\cite{PDG2020}. The differences of the results between the alternative and nominal fits are considered the systematic uncertainties for resonant parameters. Further, the systematic uncertainties associated with the signal model are investigated by using a relativistic Breit-Wigner function with the energy-dependent width. Their differences are found to be negligible. Eventually, the mass and width of the resonance are determined to be $m_{r} = 2250~\pm~25~\pm~27$ MeV$/c^{2}$ and \mbox{$\Gamma_{r} = 125~\pm~43~\pm~15$ MeV}, respectively. In addition, the $\Gamma^{ee}_{r}\cdot Br$ is $0.9\pm0.4\pm0.4$ eV or $52.9\pm17.0\pm 13.1$ eV for the two solutions, where the first and second uncertainties are statistical and systematic, respectively.

\subsection{\label{sec:leve72}Cross section of $e^{+}e^{-}\rightarrow\omega\pi\pi$}
\begin{figure}
\begin{center}
\subfigure{
\hspace{-1.0cm}
\includegraphics[scale=0.45]{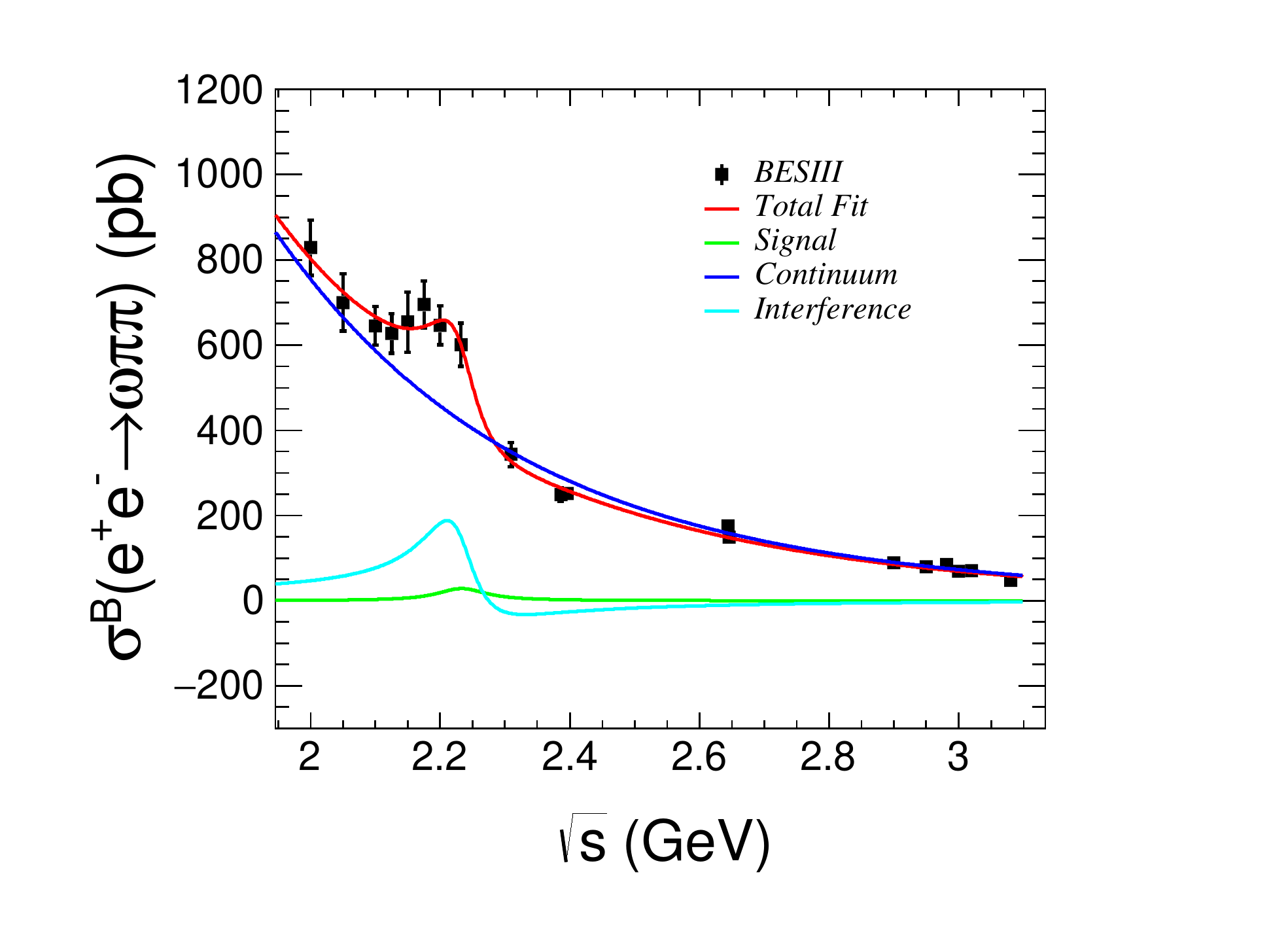}
\hspace{-1.2cm}
\includegraphics[scale=0.45]{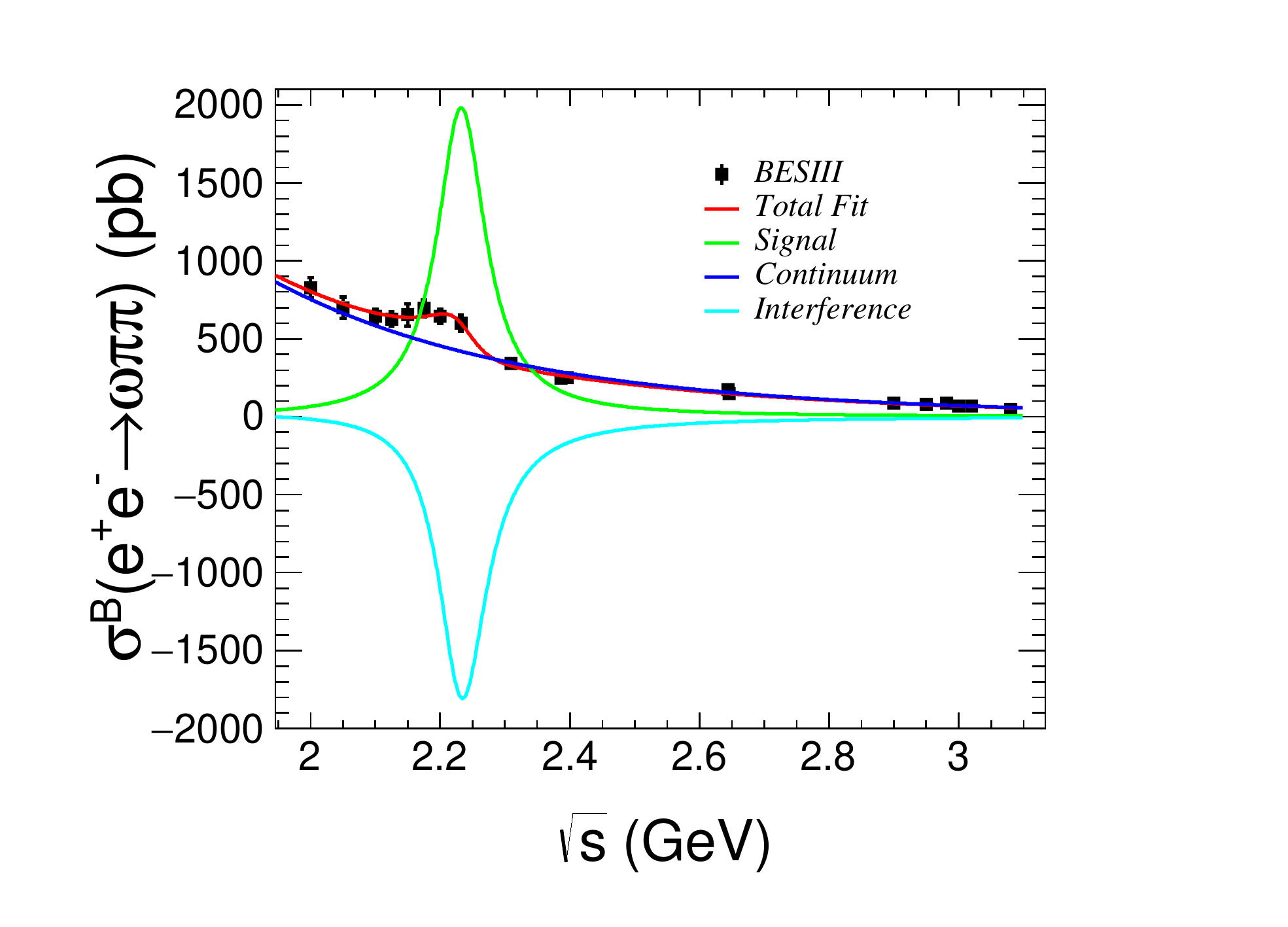}
}
\caption{\label{FitWppPlot} Fits to the cross sections of $\ee\to\omega\pi\pi$. The dots with error bars are the results combined from this work and Ref.~\cite{Bes3Wpp}, where the error bars include both statistical and systematic uncertainties. The red solid curves are the total fit results, the blue solid curves represent the continuum components, the green curves denote the resonant components, and the cyan line are the interference of the two components. The two plots are the results of Solutions I (left) and II (right), respectively.}
\end{center}
\end{figure}

 The Born cross sections for the $\ee\rightarrow\omega\pi^{+}\pi^{-}$ measured in this work are consistent with that for the $\ee\rightarrow\omega\pi^0\pi^0$ obtained in Ref.~\cite{Bes3Wpp} within a factor of 2 given their uncertainties. Therefore, the pion production in the $\ee\rightarrow\omega\pi\pi$ process complies with the SU(3)-flavor symmetry. The summed cross sections at various energy points are listed in Table~\ref{WppCrsDistr}. We fit the $e^{+}e^{-}\rightarrow\omega\pi\pi$ cross sections with the same method as in Section~\ref{sec:leve71} (see Fig.~\ref{FitWppPlot}). The resultant parameters are obtained and tabulated in \mbox{Table~\ref{FitWppResult}}. After considering the systematic uncertainty, the mass and width of X(2230) are obtained as \mbox{2232~$\pm$~19~$\pm$~27 MeV/$c^{2}$} and 91~$\pm$~53~$\pm$~20 MeV, respectively. The $\Gamma^{ee}_{r}\cdot Br$ is \mbox{0.9~$\pm$~0.5~$\pm$~0.2 eV} and 61.1~$\pm$~32.1~$\pm$~15.4 eV for the two solutions.~The statistical significance of the structure is 7.6$\sigma$.

\begin{table*}[htbp]
\begin{center}
\caption{\label{WppCrsDistr}Cross sections of $e^{+}e^{-}\rightarrow\omega\pi\pi$, where the first and second uncertainties are statistical and systematic, respectively.}
\vspace{0.3cm}
\begin{tabular}{lccc} \hline\hline
$\sqrt s$ (GeV) & Cross section~(pb) &  $\sqrt s $ & Cross section~(pb)\\ \hline
2.0000&   831.2 $\pm$ 28.2 $\pm$   36.1&  2.3960&   251.6 $\pm$  6.2 $\pm$ 10.9\\
2.0500&   697.6 $\pm$ 45.8 $\pm$   30.1&  2.6444&   176.6 $\pm$  7.2 $\pm$ 7.5\\
2.1000&   642.7 $\pm$ 23.1 $\pm$   27.3&  2.6464&   148.0 $\pm$  6.4 $\pm$ 6.5\\
2.1250&   624.8 $\pm$ 7.5 $\pm$    26.6&  2.9000&   88.40 $\pm$  2.8 $\pm$ 4.0\\
2.1500&   653.0 $\pm$ 49.0 $\pm$   32.4&  2.9500&   79.70 $\pm$  8.0 $\pm$ 4.2\\
2.1750&   698.9 $\pm$ 25.1 $\pm$   29.1&  2.9810&   85.70 $\pm$  7.4 $\pm$ 3.9\\
2.2000&   641.4 $\pm$ 21.2 $\pm$   26.5&  3.0000&   68.20 $\pm$  6.5 $\pm$ 3.7\\
2.2324&   601.2 $\pm$ 21.4 $\pm$   26.3&  3.0200&   69.70 $\pm$  6.2 $\pm$ 3.4\\
2.3096&   343.7 $\pm$ 12.9 $\pm$   16.1&  3.0800&   47.90 $\pm$  2.2 $\pm$ 2.8\\
2.3864&   249.0 $\pm$ 11.2 $\pm$   10.3&       &                            \\
\hline\hline
\end{tabular}
\end{center}
\end{table*}

\begin{table*}[htbp]
\begin{center}
\caption{\label{FitWppResult} Resonance parameters obtained in the fit to the $\ee\to \omega \pi\pi$ cross section.~The uncertainties are statistical only.}
\vspace{0.3cm}
\begin{tabular}{L{2.3cm}R{1.5cm}PR{0.7cm}R{1.5cm}PR{0.7cm}r} \hline\hline
Parameter  &\multicolumn{2}{r} {Solution I} &\multicolumn{2}{r}{Solution II}\\\hline
$m_{r}$ (MeV/$c^{2}$)&\multicolumn{4}{c}{2232~$\pm$~19}& \\
$\Gamma_{r}$ (MeV) &\multicolumn{4}{c}{93~$\pm$~53}&\\
$\Gamma^{ee}_{r}\cdot Br$(eV)&0.9&0.5&61.1 &32.1&\\
$\phi$ (rad) &2.4&0.3&$-$1.7 & 0.1&\\
$a (10^{3})$(pb$^{1/2}$)  &\multicolumn{4}{c}{1.7~$\pm$~0.2}&\\
$b$  &\multicolumn{4}{c}{4.6~$\pm$~0.1}& \\ \hline
Significance & \multicolumn{4}{c}{7.6$\sigma$} \\ \hline\hline
\end{tabular}
\end{center}
\end{table*}

We also compute the contribution of our cross section $\sigma^\text{Born}(e^{+}e^{-}\to\omega\pi\pi)$ to the hadronic contribution of $(g-2)_{\mu}$,
\begin{eqnarray}
a^{\omega\pi\pi}_{\mu}=\frac{1}{4\pi^{3}}\int_{(2.000~\textrm{GeV})^{2}}^{(3.080~\textrm{GeV})^{2}}ds^{\prime}K(s^{\prime})\sigma^\textrm{Born}_{\omega\pi\pi},
\end{eqnarray}
where $K(s^{\prime})$ is the kernel function~\cite{vp}. Our result, $a^{\omega\pi\pi}_{\mu}$=$(0.135~\pm~0.007~\pm~0.009)\times 10^{-10}$, is the first measurement for the $e^{+}e^{-}\to\omega\pi\pi$ process in this energy region.
\subsection{$X(2230)$ decays via intermediate states}
Using the measured Born cross sections via the intermediate states as tabulated in \mbox{Table \ref{ResCrs}}, we perform a simultaneous fit to the measured cross sections $\sigma_i^\text{exp}$ with $i=1,2,..,5$ for the modes $\omega f_0(500)$,~$\omega f_0(980)$, $\omega f_0(1370)$, $\omega f_2(1270)$, and $b_1(1235)\pi$, respectively. At the energy point $\sqrt s$, the cross section for the mode $i$ is described by
\begin{eqnarray}
\sigma^\text{fit}_i(\sqrt s)=|\sigma_\text{r}e^{i\phi}(\sqrt s,i)+\sigma_\text{c}(\sqrt s,i)|^2
\end{eqnarray}
 Where $\sigma_\text{r}$ and $\sigma_\text{c}$ are given by Eqs.~(\ref{SimuFitEq}) and \mbox{Eqs.~(\ref{SimuFitEq1})}, except that the phase space factor is replaced by that of the two-body decay. Considering the error correlations among these five modes, the $\chi^2$ function is minimized, which is defined by
\begin{eqnarray}
\chi^2 &=& \sum_{\sqrt s} \Sigma^T V^{-1} \Sigma,\text{~with~} \nonumber\\
\Sigma^T&=&(\sigma_1^\text{exp}-\sigma_1^\text{fit},\sigma_2^\text{exp}-\sigma_2^\text{fit},\sigma_3^\text{exp}-\sigma_3^\text{fit},\sigma_4^\text{exp}-\sigma_4^\text{fit},\sigma_5^\text{exp}-\sigma_5^\text{fit}),
\end{eqnarray}
where $V$ is a covariance matrix with element $V_{ij}=\rho_{ij}\sigma_i\sigma_j$ and $\rho_{ij}$ is a correlation coefficient and determined by the errors of interference between modes $i$ and $j$. We take $\rho_{ij}=1$ if $i=j$. The sum over $\sqrt s$ runs over all energy points involved in the fit.

\begin{figure*}[htbp]
\begin{center}
\includegraphics[width=\textwidth]{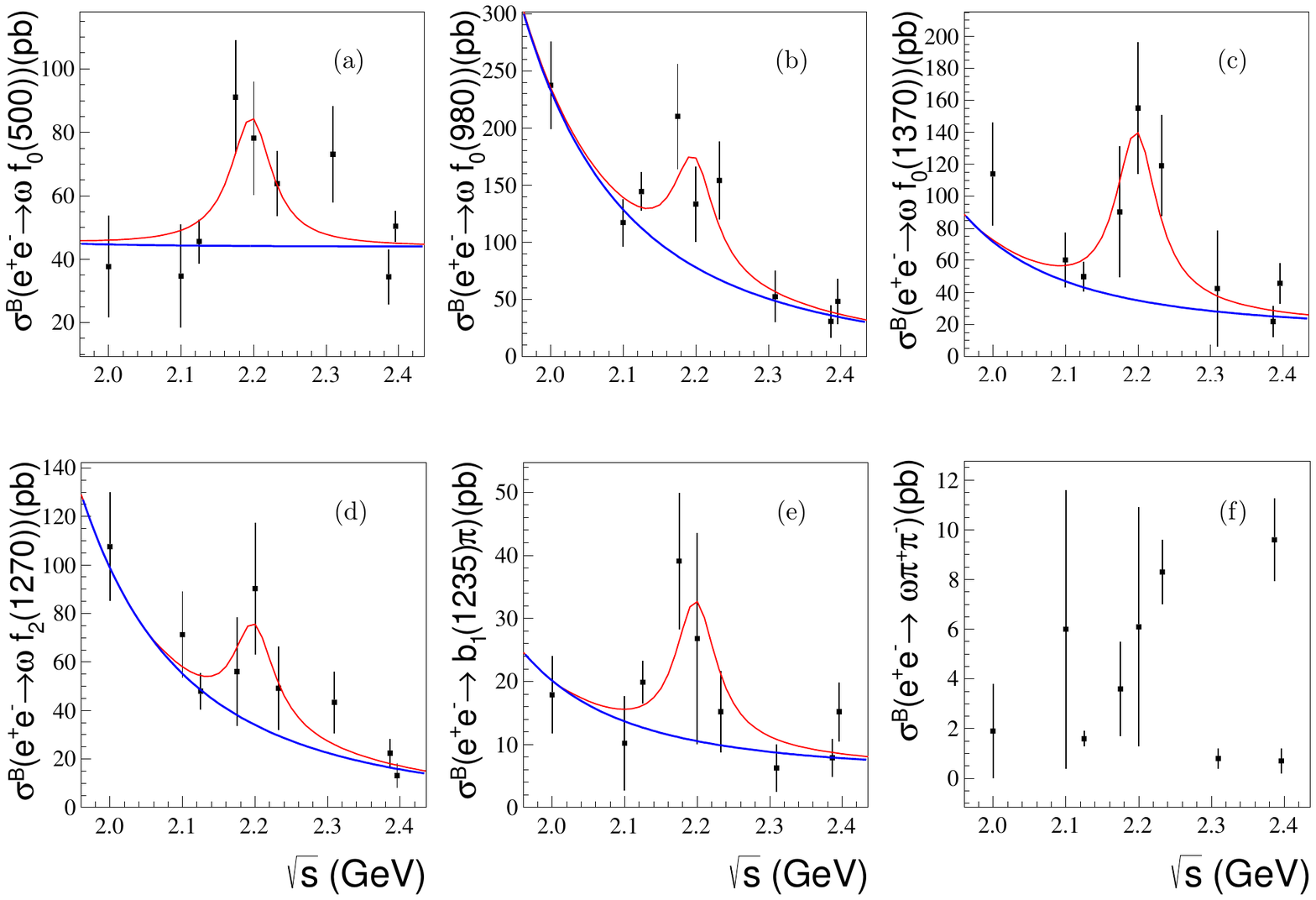}
\caption{Fits to the intermediate processes of $\ee\rightarrow$ $\omega f_{0}(500)$ (a), $\omega f_{0}(980)$ (b), $\omega f_{0}(1370)$ (c), $\omega f_{2}(1270)$ (d), $b_{1}(1235)\pi$ (e). Points with error bars are the measured Born cross sections, the red curves are the total fit results, and the blue curves denote the contributions of non-resonant processes. Plot (f) displays the contribution from the phase space process $e^{+}e^{-}\rightarrow \omega \pi^{+}\pi^{-}$.}\label{fig:C}
\end{center}
\end{figure*}

Figure~\ref{fig:C} shows results of the simultaneous fit to these five intermediate modes. In the fit, the state $X(2230)$ is assumed as a vector meson, and its mass and width is \mbox{2200~$\pm$~11~$\pm$~17 $\text{MeV}/c^{2}$} and 74~$\pm$~20~$\pm$~24 MeV, respectively. Here, the systematic uncertainties of the mass and width are determined by using the same method as in Section~\ref{sec:leve71}. The statistical significance of this structure is about 7.9$\sigma$. The resultant values of $\Gamma^{ee}_{r}\cdot Br$ are tabulated in Table~\ref{intcs}. Here, $\Gamma^{ee}_{r}$ is the partial width of the $X(2230)$ decaying to $\ee$, and $Br$ is the branching fraction for the $X(2230)$ decaying to the final state $\omega\pipi$. Two solutions are found with equally good fit quality. 
No structure around 2.25 GeV is observed in the non-resonant process, as shown in Fig.~\ref{fig:C}(f), indicating that the vector meson favors the intermediate decay.

\begin{table}[htbp]
\begin{center}
\caption{\label{intcs} $\Gamma^{ee}_{r}\cdot Br$ values from the simultaneous fit to the cross sections for the five intermediate modes, where the uncertainties are statistical only.}
\vspace{0.3cm}
\begin{tabular}{lcccc} \hline\hline
 &{Solution I} & {Solution II}&&\\ \hline
{$\ee \to f$} & $\Gamma^{ee}_{r}\cdot Br$ (eV) & $\Gamma^{ee}_{r}\cdot Br$ (eV)\\ \hline
$\omega f_0(500)$ &2.9~$\pm$~2.5&1.2~$\pm$~0.5\\
$\omega f_0(980)$ &5.9~$\pm$~5.0&3.2~$\pm$~1.0\\
$\omega f_0(1370)$ &6.4~$\pm$~5.4&3.1~$\pm~$1.0\\
$\omega f_2(1270)$ &2.5~$\pm$~2.1&1.1~$\pm$~0.6 \\
$b_1(1235)\pi$    &1.1~$\pm$~1.0&0.2~$\pm$~0.2 \\
\hline\hline
\end{tabular}
\end{center}
\end{table}

\section{Conclusion and discussion}
In summary, the Born cross sections of the $e^{+}e^{-}\rightarrow\omega\pi^{+}\pi^{-}$ process were measured using 647 pb$^{-1}$ data samples collected with the BESIII detector at 19 c.m. energies from \mbox{2.000 GeV} to 3.080 GeV. The precision of the measured cross sections are improved by a factor of 3 with respect to the previous measurement~\cite{BABAR1}. The resonance $X(2230)$ is observed with a significance of 7.6 $\sigma$ and thereby confirms the structure near 2.25 GeV observed by BaBar~\cite{BABAR1} and BESIII~\cite{Bes3Wpp}. The energy dependence of the total cross section for $e^{+}e^{-}\to \omega\pi^{+}\pi^{-}$ and $\omega\pi^{0}\pi^{0}$ is shown in Fig.~\ref{FitWppPlot}. From the fit to the $e^{+}e^{-}\rightarrow\omega\pi\pi$ cross sections, we obtained the mass and width of $X(2230)$ as 2232~$\pm$~19~$\pm$~27~MeV/$c^{2}$ and 93~$\pm$~53~$\pm$~20 MeV, respectively. The fitted $\Gamma^{ee}_{r}\cdot Br$ values are $0.9~\pm~0.5~\pm0.2$ eV and \mbox{$61.1~\pm~32.1~\pm~15.4$ eV} depending on the interference pattern. We also fit the Born cross section of the intermediate processes, such as $e^{+}e^{-}\rightarrow \omega f_{0}(500)$, $\omega f_{0}(980)$, $\omega f_{0}(1370)$, $\omega f_{2}(1270)$, and $b_{1}(1235)\pi$, with the results presented in Fig.~\ref{fig:C}. Precision measurement of the cross sections of these intermediate subprocesses will help to reveal the properties and nature of the structure. For the process $e^{+}e^{-}\rightarrow\omega\pi\pi$, we computed its hadronic contribution of the cross section to $(g-2)_{\mu}$ and found that its contribution is small in the energy region between 2.000 GeV and 3.080 GeV.
\section{Acknowledgments}
 The BESIII collaboration thanks the staff of BEPCII and the IHEP computing center for their strong support. This work is partially supported by the National Key Research and
Development Program of China (under Contract Nos. 2020YFA0406400 and 2020YFA0406300), National Natural Science Foundation of China (NSFC; under Contract Nos. 11875262, 12175244, 12175244, 11835012, 11625523, 11635010, 11735014, 11822506, 11835\\-012, 11935015, 11935016, 11935018, 11625523, 11605196, 11605198, 11705192, 12035013, 11961141012, 11950410506, and 12061131003), the Chinese Academy of Sciences (CAS) Large-Scale Scientific Facility Program, Joint Large-Scale Scientific Facility Funds of the NSFC and CAS (under Contract Nos. U2032110, U1732263, U1832207, U1832103, and U2032111), CAS Key Research Program of Frontier Sciences (under Contract Nos. QYZDJ-SSW-SLH003 and QYZDJ-SSW-SLH040), 100 Talents Program of CAS, INPAC and Shanghai Key Laboratory for Particle Physics and Cosmology, ERC (under Contract No. 758462), German Research Foundation DFG (under Contract No. 443159800, Collaborative Research Center CRC 1044, FOR 2359, FOR 2359, GRK 214), Istituto Nazionale di Fisica Nucleare (Italy), Ministry of Development of Turkey (under Contract No. DPT2006K-120470), National Science and Technology fund, Olle Engkvist Foundation (under Contract No. 200-0605), STFC (United Kingdom), The Knut and Alice Wallenberg Foundation (Sweden; under Contract No. 2016.0157), The Royal Society (UK, under Contract Nos. DH140054 and DH160214), The Swedish Research Council, and U. S. Department of Energy (under Contract Nos. DE-FG02-05ER41374 and DE-SC-0012069).

\vspace{0.5cm}
{\bf BESIII collaboration}\\

M.~Ablikim$^{1}$, M.~N.~Achasov$^{10,b}$, P.~Adlarson$^{68}$, S. ~Ahmed$^{14}$, M.~Albrecht$^{4}$, R.~Aliberti$^{28}$, A.~Amoroso$^{67A,67C}$, M.~R.~An$^{32}$, Q.~An$^{64,50}$, X.~H.~Bai$^{58}$
% ...
\\
\vspace{0.2cm}\\
{\it
$^{1}$ Institute of High Energy Physics, Beijing 100049, People's Republic of China\\
% ...
$^{4}$ Bochum Ruhr-University, D-44780 Bochum, Germany\\
% ...
$^{10}$ G.I. Budker Institute of Nuclear Physics SB RAS (BINP), Novosibirsk 630090, Russia\\
% ...
$^{14}$ Helmholtz Institute Mainz, Staudinger Weg 18, D-55099 Mainz, Germany\\
% ...
$^{28}$ Johannes Gutenberg University of Mainz, Johann-Joachim-Becher-Weg 45, D-55099 Mainz, Germany\\
% ...
$^{32}$ Liaoning Normal University, Dalian 116029, People's Republic of China\\
% ...
$^{50}$ State Key Laboratory of Particle Detection and Electronics, Beijing 100049, Hefei 230026, People's Republic of China\\
% ...
$^{58}$ University of Jinan, Jinan 250022, People's Republic of China\\
% ...
$^{64}$ University of Science and Technology of China, Hefei 230026, People's Republic of China\\
% ...
$^{67}$ University of Turin and INFN, (A)University of Turin, I-10125, Turin, Italy; (B)University of Eastern Piedmont, I-15121, Alessandria, Italy; (C)INFN, I-10125, Turin, Italy\\
$^{68}$ Uppsala University, Box 516, SE-75120 Uppsala, Sweden\\
% ...
$^{b}$ Also at the Novosibirsk State University, Novosibirsk, 630090, Russia\\
% ...
}

\end{document}